%% file: main.tex
\documentclass[11pt]{article}

\pdfoutput=1

\usepackage{amsthm}
\usepackage{graphicx} 
\usepackage{array} 

\usepackage{amsmath, amssymb, amsfonts, verbatim}
\usepackage{hyphenat,epsfig,subcaption,multirow}
\usepackage[font=small,labelfont=bf]{caption}
\usepackage{nicefrac}
\usepackage{paralist}

\usepackage[usenames,dvipsnames]{xcolor}
\usepackage[ruled]{algorithm2e}

\DeclareFontFamily{U}{mathx}{\hyphenchar\font45}
\DeclareFontShape{U}{mathx}{m}{n}{
      <5> <6> <7> <8> <9> <10>
      <10.95> <12> <14.4> <17.28> <20.74> <24.88>
      mathx10
      }{}
\DeclareSymbolFont{mathx}{U}{mathx}{m}{n}
\DeclareMathSymbol{\bigtimes}{1}{mathx}{"91}

\usepackage{tcolorbox}
\tcbuselibrary{skins,breakable}
\tcbset{enhanced jigsaw}

\usepackage[normalem]{ulem}
\usepackage[compact]{titlesec}

\definecolor{DarkRed}{rgb}{0.5,0.1,0.1}
\definecolor{DarkBlue}{rgb}{0.1,0.1,0.5}

\usepackage{nameref}
\definecolor{ForestGreen}{rgb}{0.1333,0.5451,0.1333}
\definecolor{Red}{rgb}{0.9,0,0}
\usepackage[linktocpage=true,
	pagebackref=true,colorlinks,
	linkcolor=DarkRed,citecolor=ForestGreen,
	bookmarks,bookmarksopen,bookmarksnumbered]
	{hyperref}
\usepackage[noabbrev,nameinlink]{cleveref}
\crefname{property}{property}{Property}
\creflabelformat{property}{(#1)#2#3}
\crefname{equation}{eq}{Eq}
\creflabelformat{equation}{(#1)#2#3}

\usepackage{bm}
\usepackage{url}
\usepackage{xspace}
\usepackage[mathscr]{euscript}

\usepackage{tikz}
\usetikzlibrary{arrows}
\usetikzlibrary{arrows.meta}
\usetikzlibrary{shapes}
\usetikzlibrary{backgrounds}
\usetikzlibrary{positioning}
\usetikzlibrary{decorations.markings}
\usetikzlibrary{decorations.pathreplacing} 
\usetikzlibrary{patterns}
\usetikzlibrary{calc}
\usetikzlibrary{fit}
\usetikzlibrary{decorations}

\usepackage[framemethod=TikZ]{mdframed}

\usepackage[noend]{algpseudocode}
\makeatletter
\def\BState{\State\hskip-\ALG@thistlm}
\makeatother

\usepackage{cite}
\usepackage{enumitem}
\setlist[itemize]{leftmargin=20pt}
\setlist[enumerate]{leftmargin=20pt}

\usepackage[margin=1in]{geometry}

\usepackage{thmtools}
\usepackage{thm-restate}

\newtheorem{theorem}{Theorem}
\newtheorem{lemma}{Lemma}[section]
\newtheorem{proposition}[lemma]{Proposition}
\newtheorem{corollary}[lemma]{Corollary}
\newtheorem{claim}[lemma]{Claim}

\newtheorem{definition}[lemma]{Definition}
\newtheorem{problem}{Problem}

\newtheorem*{claim*}{Claim}
\newtheorem*{assumption*}{Assumption}
\newtheorem*{proposition*}{Proposition}
\newtheorem*{lemma*}{Lemma}
\newtheorem*{corollary*}{Corollary}

\newtheorem*{theorem*}{Theorem}

\crefname{lemma}{Lemma}{Lemmas}
\crefname{claim}{claim}{claims}
\crefname{property}{Property}{Properties}
\crefname{invariant}{Invariant}{Invariants}

\newtheorem{mdresult}{Result}
\newenvironment{result}{\begin{mdframed}[backgroundcolor=lightgray!40,topline=false,rightline=false,leftline=false,bottomline=false,innertopmargin=2pt]\begin{mdresult}}{\end{mdresult}\end{mdframed}}

\newtheorem{remark}[lemma]{Remark}

\theoremstyle{definition}

\newtheorem*{mdproblem*}{Problem}
\newenvironment{Problem*}{\begin{mdframed}[hidealllines=false,innerleftmargin=10pt,backgroundcolor=gray!10,innertopmargin=5pt,innerbottommargin=5pt,roundcorner=10pt]\begin{mdproblem*}}{\end{mdproblem*}\end{mdframed}}
\newtheorem{mddefinition}[lemma]{Definition}

\newtheorem*{mddefinition*}{Definition}
\newenvironment{Definition*}{\begin{mdframed}[hidealllines=false,innerleftmargin=10pt,backgroundcolor=white!10,innertopmargin=5pt,innerbottommargin=5pt,roundcorner=10pt]\begin{mddefinition*}}{\end{mddefinition*}\end{mdframed}}
\newtheorem{mdremark}{Remark}

\newenvironment{ourbox}{\begin{mdframed}[hidealllines=false,innerleftmargin=10pt,backgroundcolor=white!10,innertopmargin=2pt,innerbottommargin=5pt,roundcorner=10pt]}{\end{mdframed}}

\newtheorem{mdalgorithm}{Algorithm}

\newenvironment{compress}{\vspace{-5pt}}{\vspace{-5pt}}

\allowdisplaybreaks

\renewcommand{\qed}{\nobreak \ifvmode \relax \else
      \ifdim\lastskip<1.5em \hskip-\lastskip
      \hskip1.5em plus0em minus0.5em \fi \nobreak
      \vrule height0.75em width0.5em depth0.25em\fi}

\newcommand{\Qed}[1]{\rlap{\qed$_{\textnormal{~~\Cref{#1}}}$}}

\renewcommand{\leq}{\leqslant}
\renewcommand{\geq}{\geqslant}
\renewcommand{\le}{\leq}
\renewcommand{\ge}{\geq}

\newif\ifnotanon
\notanontrue

\input{macros}

\title{Correlation Clustering and (De)Sparsification: \\ Graph Sketches Can Match Classical Algorithms} 
\ifnotanon{
\author{Sepehr Assadi\footnote{
Supported in part by a Sloan Research Fellowship, an NSERC Discovery Grant (RGPIN-2024-04290) and a Faculty of Math Research Chair grant. (sepehr@assadi.info). \smallskip} 
\\ {\small University of Waterloo} \and
Sanjeev Khanna\footnote{Research supported in part by NSF grants CCF-2008305, CCF-2402284 (sanjeev@cis.upenn.edu).\smallskip}
\\ {\small University of Pennsylvania} \and
Aaron Putterman\footnote{Supported in part by the Simons Investigator Awards of Madhu Sudan and Salil Vadhan and NSF Award CCF-2152413 (aputterman@g.harvard.edu).} 
\\ {\small Harvard University} 
}
}\fi

\date{}

\begin{document}
\maketitle

\pagenumbering{roman}

\begin{abstract}
    Correlation clustering is a widely-used approach for clustering large data sets based only on pairwise similarity information. In recent years, there has been a steady stream of better and better classical algorithms for approximating this problem. Meanwhile, another line of research has focused on porting the classical advances to various sublinear algorithm models, including semi-streaming, Massively Parallel Computation (MPC), and distributed computing. Yet, these latter works typically rely on ad-hoc approaches that do not necessarily keep up with advances in approximation ratios achieved by classical algorithms.
    Hence, the motivating question for our work is this: can the gains made by classical algorithms for correlation clustering be ported over to sublinear algorithms in a {\em black-box manner}? We answer this question in the affirmative by introducing the paradigm of graph de-sparsification.

    \medskip
 
    A versatile approach for designing sublinear algorithms across various models is the graph (linear) sketching. It is known that one can find a cut sparsifier of a given graph---which approximately preserves cut structures---via graph sketching, and that
    this is sufficient information-theoretically for recovering a near-optimal correlation clustering solution. However, no efficient algorithms are known for this task as the resulting cut sparsifier is necessarily a weighted graph, and correlation clustering is known to be a distinctly harder problem on weighted graphs. 

    \medskip

    Our main result is a randomized linear sketch of $\widetilde{O}(n)$ size for $n$-vertex graphs, from which one can recover with high probability an $(\alpha+o(1))$-approximate correlation clustering in polynomial time, where $\alpha$ is the best approximation ratio of any polynomial time classical algorithm for (unweighted) correlation clustering. This is proved via our new de-sparsification result: we recover in {\em polynomial-time} from some $\widetilde{O}(n)$ size linear sketch of a graph $G$, an {\em unweighted}, simple graph that approximately preserves the cut structure of $G$. In fact we show that under some mild conditions, {\em any} spectral sparsifier of a graph $G$ can be de-sparsified into an unweighted simple graph with nearly the same spectrum. We believe the de-sparsification paradigm is interesting in its own right as a way of reducing graph complexity when weighted version of a problem is harder than its unweighted version.

\medskip
    Finally, we use our techniques to get efficient algorithms for correlation clustering that match the performance of best classical algorithms, in a variety of different models, including dynamic streaming, MPC, and distributed communication models. 

\end{abstract}

\clearpage

\setcounter{tocdepth}{3}
\tableofcontents
\clearpage
\pagenumbering{arabic}
\setcounter{page}{1}

\input{intro}

\input{prelim}

\input{correlation_clustering}

\input{spectral_to_spectral}

\input{sublinear_algos}

\section{Acknowledgements}

Part of this work was conducted while the authors were visiting the Simons Institute for the Theory of Computing as part of the Sublinear Algorithms program.

\bibliographystyle{halpha-abbrv}
\bibliography{general}

\clearpage
\appendix
\part*{Appendix}

\input{appendix}

\end{document}

%% file: macros.tex
\newcommand{\Leq}[1]{\ensuremath{\underset{\textnormal{#1}}\leq}}

\newcommand{\Ot}{\ensuremath{\widetilde{O}}}
\newcommand{\eps}{\ensuremath{\varepsilon}}

\newcommand{\Bracket}[1]{\Big[#1\Big]}

\newcommand{\paren}[1]{\ensuremath{\left(#1\right)}\xspace}
\newcommand{\card}[1]{\left\vert{#1}\right\vert}

\newcommand{\IR}{\ensuremath{\mathbb{R}}}

\newcommand{\set}[1]{\ensuremath{\left\{ #1 \right\}}}
\newcommand{\poly}{\mbox{\rm poly}}

\DeclareMathOperator*{\Prob}{\ensuremath{\textnormal{Pr}}}
\renewcommand{\Pr}{\Prob}

\newenvironment{tbox}{\begin{tcolorbox}[
		enlarge top by=5pt,
		enlarge bottom by=5pt,
		 breakable,
		 boxsep=0pt,
                  left=4pt,
                  right=4pt,
                  top=10pt,
                  arc=0pt,
                  boxrule=1pt,toprule=1pt,
                  colback=white
                  ]
	}
{\end{tcolorbox}}

\newcommand{\II}{\ensuremath{\mathbb{I}}}

\newcommand{\mireal}[1][]{
  \ifx\relax#1\relax%
    \II(\mione \,; \mitwo)%
  \else%
    \II(\mione \,; \mitwo\mid #1)%
  \fi
}

%% file: intro.tex
\newcommand{\alphabest}{\alpha_{\textnormal{\textsc{best}}}}

\section{Introduction}\label{sec:intro}

Correlation clustering is a widely studied problem in theoretical computer science with applications to various areas. 
Given an undirected graph $G=(V,E)$, the goal is to cluster the vertices in a way that minimizes the cost, defined as the number of edges between the clusters and the number of non-edges\footnote{By a non-edge, we mean a pair of vertices with no edges between them in the graph.} inside the clusters. We present a new approach for solving correlation clustering via graph sketching 
with approximation guarantees that can match the performance of {\em any} polynomial time algorithm for this problem. 
This immediately leads to improved algorithms for this problem across different sublinear algorithms models for processing massive graphs.  
The core to our approach is a new problem of its own independent interest: how do we de-sparsify 
an already sparsified graph in an efficient manner? We now elaborate more on our results and their context.

\subsection{Correlation Clustering and Graph Sketches} 

Motivated by applications to processing massive graphs, there has been a rapidly growing interest in algorithms for correlation clustering 
across various \emph{sublinear} algorithms models such as semi-streaming, Massively Parallel Computation (MPC), distributed computing, and alike. 
The key challenge in these models is that the resources available to the algorithm, say, its space or communication, is much smaller 
than the input size. Thus, the algorithms often need to `compress' or `sparsify' the input graphs before they are able to solve the problem in these models.

A highly successful paradigm here, especially when it comes to flexibility and portability across different models, is \emph{graph sketching} pioneered by~\cite{AhnGM12a}: 
one compresses the graph into a sketch through a small number of linear measurements (say, of its adjacency or Laplacian matrix) and then solve the original problem given only this sketch (see~\Cref{def:graph-sketch}). 
Graph sketching has been quite successful for various graph problems including edge connectivity~\cite{AhnGM12a}, vertex connectivity~\cite{AssadiS23}, cut sparsification~\cite{AhnGM12b}, spectral sparsification~\cite{KapralovLMMS14}, densest subgraph~\cite{McGregorTVV15}, maximum matchings~\cite{AssadiKLY16}, and subgraph counting~\cite{AhnGM12b}, among many others.  This leads to the following natural question:
 
\begin{compress}
\begin{quote}
	\emph{Can we approximate correlation clustering via graph sketching?} 
\end{quote}
\end{compress}

In some sense, this question was already settled in~\cite{BehnezhadCMT23} (building on~\cite{AhnCGMW15}): it turns out the cost of any clustering
can be specified as a sum of cut sizes of the clusters plus some normalization; as such, to preserve (near-)optimal correlation clusterings, it suffices to preserve cut values of the graph in the sketch. 
But this latter task is precisely the goal of \emph{cut sparsifiers}~\cite{BenczurK96}, which are weighted 
subgraphs of the input with only $\Ot(n/\eps^2)$ edges that preserve the value of every cut to within a $(1\pm \eps)$ factor, 
and already admit efficient graph sketches~\cite{AhnGM12b}. Thus, we can also find $(1+\eps)$-approximate correlation clusterings using graph sketching. 

There is a however a serious caveat with this approach: while information-theoretically we can recover a near-optimal solution from the sparsifier, we do not know how to do this in polynomial-time. 
In particular, recovering any solution from the sparsifier essentially amounts to solving a \emph{weighted} version of correlation clustering (given that sparsification necessarily generates a weighted graph in general), which currently only admits 
an $O(\log{n})$-approximation~\cite{DemaineEFI06}\footnote{In fact, it is shown by~\cite{DemaineEFI06} that weighted correlation clustering is equivalent to the minimum multicut problem and is thus difficult to approximate better than a $\Theta(\log{n})$ factor, and does not admit any constant factor approximation under the Unique Games Conjecture.}. 
Thus, when it comes to polynomial time algorithms, the above approach appears to hit a dead end. 

As a result of the above shortcoming, recent work has come up with different graph sketches, or more often even entirely different techniques, 
for solving correlation clustering in this context; these results mostly collect ``enough'' information from the graph through the compression so as to simulate a \emph{specific} classical algorithm (often, the pivot algorithm of~\cite{AilonCN08} but also recently more improved combinatorial algorithms in~\cite{CohenAddadLPTY24}). These approaches then lead to a host of different sublinear algorithms for this problem with different guarantees across many of these models; again, see~\Cref{sec:related} for a brief summary. 
However, this means that these techniques do not necessarily keep up with the improvements on classical algorithms on this problem---which have seen many exciting developments just recently; see~\Cref{sec:related}---%
and rely on an ad hoc approach each time for porting these improvements to sublinear algorithms as well. Thus, we can ask
a more nuanced version of our original question: 

\begin{compress}
\begin{quote}
	\emph{Can we approximate correlation clustering via graph sketching in polynomial time, matching the approximation ratio of best polynomial-time classical algorithms?}
\end{quote}
\end{compress}
\noindent
We show that the answer to this question is indeed \emph{yes}. Let $\alphabest$ denote the best approximation ratio possible for correlation clustering in polynomial time (via classical algorithms), which satisfies
\begin{align}
	1.042 \Leq{\cite{CaoCLLNV24}} \alphabest \Leq{\cite{CaoCLLNV24}} 1.437 \label{eq:alpha-best}. 
\end{align} 
due to the APX-hardness established in~\cite{CharikarGW03, CaoCLLNV24} and the recent approximation algorithm of~\cite{CaoCLLNV24}. 

\begin{result}\label{res:sketch}
	For any $n$-vertex graph $G$, there is a randomized linear sketch of $\Ot(n)$ size from which one can recover with high probability an $(\alphabest+o(1))$-approximate correlation clustering of $G$ in polynomial time. 
\end{result}

We establish this result by introducing a new direction of research, termed \emph{de-sparsification}, and then use it for our particular application to correlation clustering.

\subsection{A New Question: Graph Simplification via de-sparsification?} 

Let us revisit the approaches of~\cite{AhnCGMW15,BehnezhadCMT23} that designed $(1+\eps)$-approximate graph sketches for correlation clustering (information-theoretically) via \emph{weighted} cut sparsifiers. As stated earlier, the weights in the sparsifier forces us to solve a weighted correlation clustering instance which is a much harder version of the problem than the unweighted one. But what if these instances can be made unweighted\footnote{Here, and throughout the paper, we use unweighted to also mean \emph{simple} (i.e., that there are no multi-edges permitted).} while
preserving their correlation clustering structure? This will then allow us to run 
any approximation algorithm for unweighted correlation clustering on this instance and recover essentially the same approximation guarantee on the original graph.  

To address this, we ask a general question that is entirely independent of correlation clustering:  

\begin{compress}
\begin{quote}
	\emph{Can we efficiently de-sparsify a weighted cut sparsifier $H$ to an unweighted, simple (but not necessarily sparse) graph $G$ while (nearly) preserving the value of every cut? }
\end{quote}
\end{compress}
Two important remarks are in order: $(a)$ firstly, the cut structure of weighted graphs is more general than unweighted ones, and thus in general, we cannot hope for approximating an arbitrary weighted graph with a (simple) unweighted graph; however, in our case, the weighted graph $H$ is a sparsifier of an unweighted graph and thus certainly can be approximated by an unweighted graph; $(b)$ secondly, information-theoretically it is impossible to recover the \emph{original} graph
from its weighted sparsifier due to the many-to-one nature of the sparsification; but here our goal is to find \emph{some} graph that approximates the cut structure of $H$ (and by extension the original graph), and hence, we do not run into this information-theoretic barrier. Note that there has been prior work on a topic called ``densification'' \cite{HST12}. This line of work seeks to understand when weighted graphs \emph{admit} dense sparsifiers. In our case, the question is not an existential one, but an algorithmic one. I.e., how do we \emph{efficiently recover} these denser sparsifiers. 

Unfortunately, we do not know a definitive answer to the above question in \emph{this} formulation. It seems likely that the answer is \emph{no} as it is even NP-hard to determine if a given weighted graph $H$ is a cut sparsifier of a given graph $G$ or not\footnote{We suspect this result is folklore but do not know a reference for it and thus we prove it in \Cref{sec:NP}.}. This suggests that preserving only the structure of the cuts may not allow for an efficient recovery/de-sparsification. But this then naturally suggests an alternative direction: what if we instead use \emph{spectral sparsifiers}~\cite{SpeilmanT08} that are generally known to be a robust strengthening of cut sparsifiers?\footnote{Formally, using $L_G$ and $L_H$ to denote the Laplacian matrix of $G$ and $H$, respectively, a cut sparsifier $H$ of $G$ satisfies $x^\top \cdot L_H \cdot x = (1 \pm \eps) \cdot x^\top L_G \cdot x$ for all $x \in \set{0,1}^{n}$ whereas a spectral sparsifier satisfies the same for all $x \in \IR^n$. Note that
the evidence earlier no longer holds as checking if $H$ is a spectral sparsifier of $G$ boils down to checking if all singular values of $L_H$ and $L_G$ are within $(1\pm \eps)$ factor of each other, which can be easily done in polynomial time.}

Leveraging spectral sparsifiers, we obtain the following general de-sparsification result:

\begin{result}\label{res:de-sparsify-spectral}
For any $n$-vertex unweighted graph $G$ and any $\eps \in (0,1)$, there is a randomized linear sketch of $\Ot(n/\eps^2)$ size from which one can recover in polynomial-time with high probability another $n$-vertex unweighted graph $\tilde G$ with the same number of edges as $G$ such that $\tilde G$ is a $(1 \pm \eps)$-spectral sparsifier\footnote{We note that using the term `sparsifier' might be an abuse of notation here: graph $\tilde G$ has the same exact number of edges as $G$ and thus is not sparser than $G$ in any way (nor is a subgraph of $G$). We only use the term spectral sparsifier, here and throughout the paper, to mean that its spectrum is nearly the same as $G$.} of $G$.
\end{result}

As we will show later,~\Cref{res:de-sparsify-spectral} turns out to be sufficient to obtain~\Cref{res:sketch} by simply setting $\eps = o(1)$, and then running any $\alpha$-approximation correlation clustering algorithm on the graph $\tilde G$.

\Cref{res:de-sparsify-spectral} gives efficient de-sparsification for sparsifiers obtained by a particular linear sketching scheme. Specifically, it relies on the linear sketches for spectral sparsification given in~\cite{KapralovLMMS14} that faithfully implement effective resistance-based sampling. One may ask the question if in fact any {\em arbitrary} spectral sparsifier can be efficiently de-sparsified, no matter how it was created. We show that the answer to this question is in the affirmative as well, assuming some mild conditions on the underlying graph.

\begin{result}\label{res:de-sparsify-general}
For any $\eps \in (0,1)$ and $n$-vertex unweighted graph $G$, there is a randomized polynomial-time algorithm that given any $(1 \pm \eps)$-spectral sparsifier $H$ of $G$, recovers with high probability 
\begin{itemize}
\item[(a)] an $n$-vertex unweighted graph $\tilde G$ with the same number of edges as $G$ such that $\tilde G$ is a $(1 \pm 2\eps)$-cut sparsifier of $G$, provided the minimum cut in $G$ is $\Omega(\log n/\eps^2)$.

\item[(b)]
an $n$-vertex unweighted graph $\tilde G$ with the same number of edges as $G$ such that $\tilde G$ is a $(1 \pm 2\eps)$-spectral sparsifier of $G$, provided  maximum effective resistance in $G$ is $O(\eps^2/\log n)$.
\end{itemize}
\end{result}

 We believe the de-sparsification question posed in this paper to be of its own independent interest specifically from the view point of \emph{graph simplification}: while traditionally one often considers simplifying a graph as making it sparser, it is also quite natural to simplify a graph by making it \emph{unweighted} even at the cost of increasing its density (given that many problems are easier to solve on unweighted graphs such as correlation clustering considered here). Such simplification questions have also recently been considered for other graph problems in entirely different contexts, e.g., for preserving shortest path structures without using very large weights~\cite{BernsteinBW24}, or for approximating weighted matching via unweighted matching algorithms~\cite{BernsteinDL21,BernsteinCDZST24}.

\subsection{Implication to Sublinear Algorithms}\label{sec:sublinear}

Finally, we can use our efficient graph sketches for correlation clustering in~\Cref{res:sketch} to obtain new sublinear algorithms for this problem across a variety of different models, that can achieve approximation ratios nearly matching $\alphabest$ defined in~\Cref{eq:alpha-best}. 

Our first algorithm is in the distributed communication model, studied for various clustering problems in~\cite{ChenSWZ16,AwasthiBBWW19,ZhuZLHB19} (although we are not aware of prior work on correlation clustering here). In this model, 
the input graph $G=(V,E)$ is edge-partitioned across $k$ machines plus a coordinator that receives no input. The machines and the coordinator can communicate in a distributed point-to-point manner. The goal is to limit the total communication while allowing the coordinator to output a correlation clustering of the entire input. 

\begin{corollary}\label{cor:distributed}
    There is a polynomial-time randomized algorithm for correlation clustering in the distributed communication model with $k$ machines that uses $\Ot(nk)$ communication in total, and with high probability, achieves an $(\alphabest+o(1))$-approximation. 
\end{corollary}

The second algorithm is in the Massively Parallel Computation (MPC) model~\cite{KarloffSV10,BeameKS17}. Here, the input graph $G = (V,E)$ is edge-partitioned across multiple machines. Computation happens in synchronous rounds wherein each machine can send and receive $\Ot(n)$-size messages. After the last round, one designated machine outputs a solution to the problem. 

\begin{corollary}\label{cor:mpc}
	There is a polynomial-time randomized algorithm for correlation clustering in the MPC model that uses $O(1)$ rounds and $\Ot(n)$-size messages per-machine, and with high probability, achieves an $(\alphabest+o(1))$-approximation. 
\end{corollary}

\Cref{cor:mpc} improves upon the 1.87-approximation MPC algorithm of~\cite{CohenAddadLPTY24} (given~\Cref{eq:alpha-best}), although we note that the algorithm of~\cite{CohenAddadLPTY24} runs in $O(1)$ rounds
even when memory per machine is $n^{\delta}$ for any constant $\delta \in (0,1)$. But importantly, our algorithm in~\Cref{cor:mpc} has the benefit of automatically improving in future using any other advances on classical algorithms for correlation clustering.  

We can also implement our algorithm in the dynamic streaming model. Here, the input graph $G=(V,E)$ is presented to the algorithm as a stream of edge insertions and deletions, and the algorithm can make a 
single pass (or a few passes) over this stream and should output the answer to the problem on the graph $G$ at the end. 

\begin{corollary}\label{cor:dynamicStreaming}
	There is a polynomial-time randomized streaming algorithm for correlation clustering that uses $\Ot(n)$ memory when making a single pass over a dynamic stream, and with high probability, achieves an 
 $(\alphabest+o(1))$-approximation. 
\end{corollary}
Again, our result improves upon the prior $3$-approximation algorithm of~\cite{CambusKLPU24} in dynamic streams and 1.84-approximation algorithm of~\cite{CohenAddadLPTY24} in insertion-only streams. 

Finally, our approach also have an interesting consequence to insertion-only streams using non-sketching techniques (in particular using~\Cref{res:de-sparsify-general} and not~\Cref{res:de-sparsify-spectral} used in our other algorithms): it provides the first polynomial time algorithm that processes the stream \emph{deterministically} and uses randomness only at the end of the stream. This guarantee in particular satisfies the notion of adversarially-robust streaming algorithms~\cite{Ben-EliezerJWY22} in the strongest possible sense as it works even against an adversary that sees its internal state; see also~\cite{ChakrabartiGS22}. 

\begin{corollary}\label{cor:insertionStreaming}
	There is a polynomial-time streaming algorithm for correlation clustering that uses $\Ot(n)$ memory to deterministically build a data structure $D$ using a single pass over an insertion-only stream, and only at the end, uses randomization to, with high probability, recover from $D$ an $(\alphabest+o(1))$-approximation for correlation clustering.
\end{corollary}

Before moving on, an important remark about our sublinear algorithms is in order. 
\begin{remark}
\emph{Our approach inherently bounds the size of the sketch it computes and not the post-processing algorithms (given we have no control over the space-complexity of the best classical algorithm we run at the end beside it being polynomial). For our distributed algorithms, this is inconsequential. In the MPC model, this means the in- and out-communication by each machine will be bounded by $\Ot(n)$ (but not the internal memory) which is inline with the original definitions in~\cite{BeameKS17} (see also~\cite{RoughgardenVW18}) that allow for any complex operations to be done on each machine. For the streaming algorithms, this means that the memory of the algorithm \emph{during} the stream is bounded by $\Ot(n)$ but after the stream finishes, to recover the solution, the space used by the algorithm may become larger. We note that to our knowledge, \underline{all} existing streaming \emph{lower bounds} only bound the space of the algorithm during the stream\footnote{Specifically, the techniques in communication complexity and branching programs used for proving streaming lower bounds are inherently oblivious to the post-processing space of the algorithm.}.
} 
\end{remark}

\subsection{Our Techniques}\label{sec:techniques}

Our approach in establishing~\Cref{res:de-sparsify-general} consists of two  steps: (1) recovering a \emph{fractional} sparsifier, namely, a graph with all edge-weights in $[0,1]$, from the given spectral sparsifier $H$ of $G$, and then, (2) \emph{rounding} this fractional sparsifier into a simple unweighted graph to obtain the graph $\tilde{G}$. We implement the first step (for both parts of this result) by formulating the problem as a convex program and devising a separation oracle to run Ellipsoid algorithm on this program (the separation oracle crucially relies on $H$ being a spectral sparsifier, as the oracle for cut sparsifiers is solving an NP-hard problem in general). The second part is done via a randomized rounding approach---which, additionally ensures the number of sampled edges \emph{exactly} matches the original graph---but requires different analysis for each part: a union bound approach using Karger's cut-bounding bound (see~\Cref{prop:cut-counting}) relying on the assumption that minimum cut is not too small, or, following the standard effective resistance sampling approach (see~\Cref{prop:SS11}) for constructing spectral sparsifiers, using the assumption that effective resistances are not too large. 

To obtain our sketch in~\Cref{res:de-sparsify-spectral} from~\Cref{res:de-sparsify-general}, we first use a sketch due to~\cite{AhnGM12a, KPS24d} that identifies $\tilde O(n)$ edges, whose removal partitions the graphs into subgraphs with large enough minimum cut as required by~\Cref{res:de-sparsify-general}; in parallel, we also use a sketch by~\cite{KapralovLMMS14} for spectral sparsification, and use linearity of these sketches to recover a sketch for each of these large-min-cut subgraphs. A final argument then shows we can use the recovered edges plus the unweighted sparsifiers on each component obtained via~\Cref{res:de-sparsify-general} to get an unweighted sparsifier of the entire graph as well (in~\Cref{sec:sketch}, we show how the plan outlined above can recover a cut sparsifier from the sketch, which is a weaker version of~\Cref{res:de-sparsify-spectral} but is sufficient for proving~\Cref{res:sketch} for correlation clustering; we then improve this to recover a spectral sparsifier in~\Cref{sec:res-spectral}). 

Finally, to obtain~\Cref{res:sketch} from our~\Cref{res:de-sparsify-spectral}, we follow previous arguments in~\cite{AhnCGMW15,BehnezhadCMT23} that show cut sparsifiers (information-theoretically) preserve correlation clustering structure; the new part here is to ensure the problem reduces to an instance of correlation clustering on the (unweighted) sparsifier (not some rather arbitrary computation as in~\cite{AhnCGMW15,BehnezhadCMT23}), which further requires us to \emph{exactly} match the number of edges in the original graph and the unweighted sparsifier (which was an additional property obtained in~\Cref{res:de-sparsify-spectral}). 
\subsection{Related Work}\label{sec:related}

The last couple of years has witnessed a flurry of results on correlation clustering both for sublinear as well as classical algorithms. 
For instance,~\cite{CohenAddadLMNP21}
designed $O(1)$-round MPC algorithms for correlation clustering, and building on this,~\cite{AssadiW22} obtained single-pass streaming and sublinear time $O(1)$-approximation algorithms for this problem\footnote{The constants in these algorithms are quite
large, around 700 for~\cite{CohenAddadLMNP21} and more than $10000$ for~\cite{AssadiW22}.} (see also~\cite{AhnCGMW15,ChierichettiDK14} and references therein for earlier work on this problem). These results were subsequently improved in a series of work in~\cite{BehnezhadCMT22,BehnezhadCMT23,DalirrooyfardMM24,MakarychevC23,CambusKLPU24,CohenAddadLPTY24} culminating in the work of~\cite{CohenAddadLPTY24} that achieves a 1.847-approximation via single-pass streaming or sublinear time algorithms and 1.876-approximation in $O(1)$ MPC rounds (considerably simpler algorithms achieving a $(3+\eps)$-approximation were also developed in~\cite{MakarychevC23,CambusKLPU24} 
by adapting the landmark Pivot algorithm of~\cite{AilonCN08} to these models). 

Meanwhile, there has also been exciting progress on classical algorithms for correlation clustering. Early work on this problem led to $3$-approximation combinatorial and 2.5-approximation LP based algorithms for this problem~\cite{AilonCN08}, 
which was then improved to a 2.06~\cite{ChawlaMSY15}. Recently,~\cite{CohenAddadLN22} broke the $2$-approximation barrier---the integrality gap of the LP of~\cite{AilonCN08}---and achieved a 1.995-approximation which was then improved to a 1.73-approximation in~\cite{CohenAddadL0N23} and subsequently 1.437-approximation in~\cite{CaoCLLNV24}. Finally,~\cite{CohenAddadLPTY24} gave a combinatorial 1.84-approximation algorithm 
which, as stated earlier, can also be implemented in streaming and sublinear time (and with some small loss MPC) models.

%% file: prelim.tex
\newcommand{\cut}[2]{\mathrm{cut}_{#1}(#2)}
\newcommand{\mincut}[1]{\mathrm{mincut}(#1)}

\section{Preliminaries}\label{sec:prelim}

\paragraph{Notation.}\label{sec:CCnotation} Throughout, we work with undirected graphs $G=(V,E)$ and use $n$
to denote the number of vertices in $G$. For a set $S \subseteq V$, we use $G[S]$ to denote the induced subgraph of $G$ on $S$, and $\delta(S)$ to denote the edges crossing the cut induced by $S$. For a vertex $v \in V$, we use $N(v)$ and $\deg(v)=\card{N(v)}$ to denote its neighbors and its degree, respectively. 

For weighted graphs, we use $w_G: E \rightarrow \IR$ to denote the weights. We often treat unweighted graphs as weighted graphs with weight one on every edge (primarily, to avoid repeating definitions for them separately). For a cut $S \subseteq V$, we use $\cut{G}{S} := \sum_{e \in \delta(S)} w_G(e)$ to denote the weight of the edges in the cut. We use 
$\mincut{G}$ to denote the minimum cut value in $G$.
Additionally, we use $L_G$ to denote the Laplacian matrix of the graph $G$, where $(L_G)_{u,u}$ is the weighted degree of each vertex $u \in V$, and $(L_G)_{u,v} = -w_{u,v}$ for each edge $(u,v) \in E$ and $0$ otherwise. 

We say an event happens \emph{with high probability} if its probability is at least $1-1/\poly(n)$ where $n$ is the number of vertices in the underlying graph (which will be clear from the context). 

Likewise, we will often use the shorthand $a \in (1 \pm \eps)b$ to mean that $(1 - \eps)b \leq a \leq (1 + \eps)b$.

\paragraph{Correlation clustering.} For a partition $V_1, \dots V_k$, we use $E^+_G(V_1, \dots V_k)$ to denote all the edges in $G$ which are crossing between $V_1, \dots V_k$.
Likewise, we use $E^-_G(V_i)$ to denote the set of non-edges (i.e., not present edges in $G$) which are contained in $V_i$.

\begin{definition}\label{def:cc}
     Let $G = (V, E)$ be an arbitrary unweighted graph. Then, for a partition $V_1, \dots V_k$, the value of the partition under the correlation clustering objective is:
    \[
    \mathrm{CC}_G(V_1, \dots V_k) = \sum_{i \in [k]} |E^-(V_i)| + |E^+(V_1, \dots V_k)|.
    \]
    The goal in the correlation clustering problem is to find a partition that \underline{minimizes} this objective. 
\end{definition}

\subsection{Cut and Spectral Sparsifiers}
We will frequently be concerned with graph sparsifiers and specifically cut sparsifiers~\cite{BenczurK96} and spectral sparsifiers~\cite{SpeilmanT08}. We review their definitions here. 

\paragraph{Cut sparsifiers.} A basic notion of sparsification is \emph{cut sparsification} introduced by~\cite{BenczurK96}.  

\begin{definition}[\!\!\cite{BenczurK96}]\label{def:cut-sparsifer}
    Given a graph $G = (V, E)$ and $\eps \in (0,1)$, a graph $\widetilde{G}$ is said to be a $(1 \pm \eps)$ \textbf{cut sparsifier} of $G$ iff for every cut $S \subseteq V$, 
    \[
    (1 - \eps) \cdot \mathrm{cut}_G(S) \leq \mathrm{cut}_{\widetilde{G}}(S) \leq (1 + \eps) \cdot\mathrm{cut}_G(S). 
    \]
\end{definition}

We note that one often requires a cut sparsifier of a graph to be its subgraph. However, as stated in~\Cref{res:de-sparsify-spectral}, this is not the case in our paper due to our de-sparsification approach (which does not require this guarantee, nor can provide it without trivializing the problem).

 A key quantity of interest when designing cut sparsifiers is known as the \emph{strength} of an edge: 

 \begin{definition}\label{def:strength}
     Given a graph $G = (V, E)$, the \textbf{strength} of an edge $e \in E$ is defined as
     \[
     \lambda_e = \max_{S \subseteq V: e \subseteq S} \mathrm{mincut}(G[S]).
     \]
 \end{definition}

\paragraph{Spectral sparsifiers.}
A strictly stronger notion than cut sparsifiers 
are \emph{spectral sparsifiers}~\cite{SpeilmanT08}.  

\begin{definition}[\!\!\cite{SpeilmanT08}]\label{def:spectral-sparsifer}
Given a graph $G = (V, E)$, a graph $\widetilde{G}$ is considered a $(1 \pm \eps)$ \textbf{spectral sparsifier} of $G$ iff
for every vector $x \in \IR^V$, 
    \[
    (1 - \eps) \cdot x^\top L_G x \leq x^\top L_{\widetilde{G}} x \leq (1 + \eps) \cdot x^T L_G x,
    \]
    where $L_{G}$ and $L_{\widetilde{G}}$ denote the Laplacian matrix of $G$ and $\widetilde{G}$, respectively. 
\end{definition}

Similar to strength of edges defined in the context of cut sparsifiers, we have effective resistances for spectral sparsifiers. 

\begin{definition}\label{def:ER}
    For a graph $G = (V, E)$, and a pair of vertices $(u,v) \in \binom{V}{2}$, we say that the \textbf{effective resistance of $(u,v)$ in $G$} is:
    \[
    R_{\mathrm{eff}, G}(u,v) = \max_{x \in \mathbb{R}^V, x \neq 0} \frac{(x_u - x_v)^2}{x^T L_G x}.
    \]
\end{definition}

Finally, we need some additional properties from sparsifiers captured in the following definition.

\begin{definition}\label{def:totalweightpreservingcutsparsifier}
    Given a $(1 \pm \eps)$ cut/spectral sparsifier $\widetilde{G}$ of a graph $G = (V, E)$, we say that $\widetilde{G}$ is \textbf{total weight preserving} if it additionally satisfies $\sum_{e \in G} w_G(e) = \sum_{e \in \widetilde{G}} w_{\widetilde{G}}(e)$. Similarly, 
    we say $\widetilde{G}$ is \textbf{simple} iff it is an unweighted simple graph. 
\end{definition}

\subsection{Graph Sketches}\label{sec:graph-sketch}

In this work, we will frequently be concerned with designing (linear) graph sketches, introduced in the work of~\cite{AhnGM12a} (for graph problems). 

\begin{definition}\label{def:graph-sketch}
    A \textbf{linear sketch} of a graph $G$ is identified by a (possibly randomized) matrix $M$ of dimensions $s \times {{n}\choose{2}}$, chosen independently of the graph. Then, given an unweighted graph $G$ with edge incidence vector 
    $\mathbf{1}_G \in \set{0,1}^{{n}\choose{2}}$, the \textbf{sketch} of the graph is given by $M \cdot \mathbf{1}_G$. Finally, there is a \textbf{recovery algorithm} that given only the sketch and the sketching matrix, with no direct access to $G$, outputs the solution to a given problem on $G$.  
\end{definition}

The convention is that the entries in the linear sketch should be bounded in magnitude by $\mathrm{poly}(n)$, and thus the space complexity of the linear sketch is $O(s \log(n))$ bits. 

We note that even though we work with both weighted and unweighted graphs in this work, we have opted to define the sketching only for unweighted graphs given certain subtleties in the definition for weighted graphs, which will not be relevant to our work (see~\cite{ChenKL22} for more details). 

Linearity of these sketches allows one to use them in various sublinear algorithms models; we elaborate more on this in~\Cref{sec:sublinear-algs} when designing our sublinear algorithms.

%% file: correlation_clustering.tex
\section{Desparsification for Correlation Clustering: Proof of~\Cref{res:sketch}}\label{sec:sketch}

\newcommand{\cutt}{\text{cut}}
\newcommand{\R}{\mathbb{R}}
\newcommand{\Z}{\mathbb{Z}}
\newcommand{\zo}{\{0,1\}}

In this section, we prove the following theorem that formalizes~\Cref{res:sketch}. 
\begin{theorem}\label{thm:formalResult1}
    Let $\alphabest$ be the best possible approximation ratio for correlation clustering on simple graphs in polynomial time. There is a linear sketch of size $\widetilde{O}(n)$ bits, which for any simple graph $G$ on $n$ vertices can be used to recover an $(\alphabest + o(1))$ approximation to correlation clustering in $G$ in polynomial time with high probability. 
\end{theorem}

The key building block in the proof of \Cref{thm:formalResult1} is the following general de-sparsification result, which is a weaker version of \Cref{res:de-sparsify-spectral} (we opted to start with this weaker version as it suffices for our application and contains many of ideas for the full result as well).

\begin{theorem}\label{thm:sparsifierConstruction}
    There is a (randomized) linear sketch using $\widetilde{O}(n / \eps^2)$ bits of space which, for any simple graph $G$, can be used to recover with high probability a simple, $(1 \pm \eps)$ total weight preserving cut sparsifier of $G$ in polynomial time.
\end{theorem}

In the rest of this section, we first show how to use total weight preserving sparsifiers to solve correlation clustering and prove~\Cref{thm:formalResult1} using~\Cref{thm:sparsifierConstruction}. We then switch to the proof of~\Cref{thm:sparsifierConstruction} by presenting its sketch first, and then going through the two separate steps outlined in~\Cref{sec:techniques} needed for its proof. 

\subsection{Correlation Clustering from Total Weight Preserving Sparsifiers}

The following lemma motivates total weight preserving cut sparsifiers for correlation clustering. 

\begin{lemma}\label{clm:preserveCCperfectDegrees}
    Let $G$ and $H$ be graphs on the same vertex set such that $H$ is a $(1 \pm \eps)$ total weight preserving cut sparsifier of $G$. Then, for any partition $V_1, \dots V_k$ of vertices,
    \[
    \mathrm{CC}_H(V_1, \dots V_k) \in 
    (1 \pm 2\eps) \cdot \mathrm{CC}_G(V_1, \dots V_k).
    \]
\end{lemma}

We note that similar but not identical statements as~\Cref{clm:preserveCCperfectDegrees} have been used in prior work in~\cite{AhnCGMW15, BehnezhadCMT23}; as such, we postpone the proof of this lemma to~\Cref{app:omitted}.  With this lemma, we can immediately obtain \Cref{thm:formalResult1}, assuming \Cref{thm:sparsifierConstruction}.

\begin{proof}[Proof of \Cref{thm:formalResult1}]
    The linear sketch is exactly the one of \Cref{thm:sparsifierConstruction}. Let $\widetilde{G}$ denote the recovered simple graph which is a $(1 \pm \eps)$ total weight preserving cut sparsifier of $G$. By \Cref{clm:preserveCCperfectDegrees}, we see that for any clustering $V_1, \dots V_k$, 
    \[
    \mathrm{CC}_{\widetilde{G}}(V_1, \dots V_k) \in (1 \pm 2\eps) \cdot \mathrm{CC}_{G}(V_1, \dots V_k).
    \]
    So, if we let $\mathrm{OPT}(G)$ denote the minimum correlation clustering value on $G$, we know that 
    \[
    \mathrm{OPT}(\widetilde{G}) \leq (1 + 2 \eps) \cdot \mathrm{OPT}(G).
    \]

    Now, let us run any black-box $\alphabest$-approximation, polynomial time algorithm for correlation clustering on $\widetilde{G}$ in (crucially using the fact that $\widetilde{G}$ is simple). We are guaranteed that this recovers a partition $P=(V_1,\ldots,V_k)$ of vertices such that 
    \[
    \mathrm{CC}_{\widetilde{G}}(P) \leq \alpha \cdot \mathrm{OPT}(\widetilde{G}).
    \]
    Returning $P$ as the answer on $G$, by~\Cref{clm:preserveCCperfectDegrees} satisfies
    \[
     \mathrm{CC}_{G}({P}) \leq (1 + 2 \eps) \cdot \mathrm{CC}_{\widetilde{G}}({P}) \leq (1 + 2 \eps) \cdot \alpha \cdot \mathrm{OPT}(\widetilde{G}) \leq \alpha \cdot (1 + 2 \eps)^2 \cdot \mathrm{OPT}(G).
    \]

    Finally, by setting $\eps = o(1)$, the linear sketch we use requires only $\widetilde{O}(n)$ bits, yet still recovers an $(\alphabest + o(1))$-approximate solution to correlation clustering on $G$ in polynomial time. 
\end{proof}

\subsection{Building the Linear Sketch used in \Cref{thm:sparsifierConstruction}}

We now switch to proving~\Cref{thm:sparsifierConstruction} which is the main technical contribution of this section.  
We will require three distinct linear sketches for constructing our total-weight preserving sparsifier:

\begin{enumerate}
    \item First, we require a linear sketch which, for some parameter $\lambda = \Theta(\log(n) / \eps^2)$ to be chosen later, can be used to (exactly) recover all edges of strength at most $\lambda$ in the graph $G$, denoted by $S_1(G)$. This is done via the following result. 
    \begin{proposition}[cf.{~\cite{AhnGM12a},\cite[Claim 4.9]{KPS24d}}]\label{clm:KPS}
    For any given $\lambda \geq 1$, there is a linear sketch for (unweighted) graphs $G$ on $n$ vertices for recovering \underline{all} edges of strength $\leq \lambda$ with high probability in polynomial time, using $\widetilde{O}(n \lambda)$ space. 
\end{proposition}

    \item Second, we require a linear sketch which recovers a $(1 \pm \eps)$ spectral sparsifier of the graph $G$, denoted by $S_2(G)$. This is done via the following result. 
    \begin{proposition}[\!\!\cite{KLMMS14}]\label{clm:KLMMS}
    For any given $\eps \in (0,1)$, there is a linear sketch for (unweighted) graphs $G$ on $n$ vertices for recovering a $(1 \pm \eps)$ spectral sparsifier of $G$ with high probability in polynomial time, using $\widetilde{O}(n/\eps^2)$ space. 
\end{proposition}
    \item Finally, our remaining linear sketch is simply the total number of edges present in the graph. This is a deterministic linear sketch that simply tracks the size of the support of the $\binom{n}{2}$ dimensional vector describing the graph $G$, and we denote this sketch by $S_3(G)$, but will often implicitly refer to this quantity as $m$. 
\end{enumerate}

The lemma below shows how these these three linear sketches can be used to recover structural information about $G$ that will be sufficient for de-sparsification.

\begin{lemma}\label{lem:linearSketchParts}
    Given the linear sketches $S_1(G), S_2(G), S_3(G)$, one can recover:
    \begin{enumerate}
        \item a set of edges $T \subseteq G$ such that $G - T$ has minimum cut $> \lambda$,
        \item a $(1 \pm \eps)$-spectral sparsifier of the graph $G - T$, and 
        \item the total number of edges in $G - T$.
    \end{enumerate}
    Further, the space complexity of these sketches is $\widetilde{O}(n\lambda + n / \eps^2)$ bits.
\end{lemma}

\begin{proof}
    We start by using $S_1(G)$ to recover the edges of strength at most $\lambda$ in $G$, denoted by $T$. Importantly, because we recover \emph{only} these edges, the resulting graph $G - T$ has all edges of strength greater than $\lambda$. This step implicitly partitions the vertex set $V$ into $V_1, V_2, ..., V_k$ such that for $i \in [k]$, the subgraph of $G-T$ induced by $V_i$ has minimum cut greater than $\lambda$. We rely here on the basic property of edge strengths, namely, the certificate of an edge having strength greater than $\lambda$ in $G$ never uses an edge of strength at most $\lambda$. In other words, any edge in $T$ necessarily connects a vertex in some $V_i$ to a vertex in some $V_j$ for $1 \le i \neq j \le k$.

    Next, because we have recovered the set $T \subseteq G$ of edges, we can simply delete these edges from the linear sketch $S_2(G)$, yielding a linear sketch $S_2(G - T)$. Now, invoking the recovery algorithm of \Cref{clm:KLMMS}, we can recover a $(1 \pm \eps)$-spectral sparsifier for each of $G[V_1], G[V_2], ..., G[V_k]$.

    Finally, number of edges in $G - T$ is obtained by subtracting the number of edges in $T$ from $m$. 

    The space complexities of linear sketches $S_1(G)$ and $S_2(G)$ follow from \Cref{clm:KPS} and \Cref{clm:KLMMS}, respectively. The sketch $S_3(G)$ takes only $O(\log{n})$ space.
\end{proof}

The combination of $S_1, S_2, S_3$ is the entirety of the linear sketches that we will store. The rest of the complexity in our procedure is in \emph{recovering} a specific type of sparsifier. We discuss this more in the coming subsections.

\subsection{Recovering Fractional Total Weight-preserving Sparsifiers}

We now describe how given a total weight preserving spectral sparsifier $H$ of some unweighted graph $G$ with $m$ edges, we can recover in polynomial time a \emph{fractional}, total weight preserving sparsifier $\widetilde G$. We say $\widetilde G$ is fractional in the sense that every edge $e \in \widetilde G$ will have a fraction $Y_e \in [0,1]$ assigned to it, which can be seen as its weight. In the next subsection we will show how $\widetilde G$ can be rounded to an unweighted graph that is a total weight-preserving cut-sparsifier of $H$ and as such $G$ as well.

In this section, we prove the following lemma:

\begin{lemma}\label{lem:convexProgram}
    Given a (potentially weighted) graph $H$ which is promised to be a $(1 \pm \eps)$ spectral sparsifier of some unweighted graph $G$, along with the number of edges in $G$, one can recover (in polynomial time) a $(1 \pm 3\eps)$ fractional total weight preserving spectral sparsifier of the graph $G$.
\end{lemma}

\begin{proof}
    Consider the following convex program in $\R^{\binom{V}{2}}$, for which we wish to find a feasible point:
    \begin{ourbox}
    \begin{align*}
    &~~Y_e \in [0,1] \quad \forall e \in \binom{V}{2},  \\
    & \sum_{e \in \binom{V}{2}} Y_e \cdot z^\top L_e z \geq (1 - \eps)z^\top L_{H} z \quad \forall z \in \R^V: \Vert z \Vert_2 = 1, \\
    & \sum_{e \in \binom{V}{2}} Y_e \cdot z^\top L_e z \leq (1 + \eps)z^\top L_{H} z \quad \forall z \in \R^V: \Vert z \Vert_2 = 1, \\
    & \sum_{e \in \binom{V}{2}} Y_e = m.
\end{align*}
    \end{ourbox}
Here, $L_e$ is the Laplacian matrix of the $n$-vertex graph consists of only the single edge $e$. 

Each of the infinitely many constraints of this program
are linear in the variables $Y_e$ since $z^\top L_H z$ (and similar for $L_G$) are simply numbers for each fixed $z$. 
This program also has a feasible solution, as the original graph $G$ is a $(1 \pm \eps)$ spectral sparsifier of $H$ (and vice versa), with $m$ edges and weights that are $\set{0,1}$ and hence
the characteristic vector of its edges form a feasible solution to this program. 

We now show there is a polynomial time separation oracle for this program. Fix any assignment to $Y_e$'s yielding a fractional graph which we will denote by $G(Y)$, where the weight of edge $e$ is $Y_e$. We can check in polynomial time whether the total edge weights in $G(Y)$ equal $m$, that each $Y_e \in [0,1]$, and that $G(Y)$ and $H$ have the same connected components. If any of these checks fail, we have found a violated constraint. We now focus on verifying that $L_{G(Y)}$ is a $(1 \pm \eps)$ spectral sparsifier of $L_{H}$. That is, we wish to check whether 
\[
   (1- \eps) \cdot L_{H}\preceq  L_{G(Y)} \preceq (1 + \eps) \cdot L_{H},
\]
where $\preceq$ refers to the Loewner order of PSD matrices. Observe that this condition passes if and only if $G(Y)$ is a $(1 \pm \eps)$-spectral sparsifier of $H$, as 
\[
(1- \eps) \cdot L_{H}\preceq  L_{G(Y)} \preceq (1 + \eps) \cdot L_{H} \iff \forall z \in \R^V: (1- \eps) \cdot z^T L_{H} z \leq  z^T L_{G(Y)} z \leq (1 + \eps) \cdot z^T  L_{H} z,
\]
by the definition of the Loewner order.

By left and right multiplying by $L_H^{\dagger/2}$, where $L_H^{\dagger}$ is the pseudo-inverse of $L_H$ (and restricting our attention to the image of $L_H$), this is equivalent to checking whether 
 \begin{align}\label{eq:pseudoinverseCheck}
    (1 - \eps) \cdot I_{\mathrm{Im}(L_H)} \preceq L_{H}^{\dagger/2} \cdot L_{G(Y)}\cdot L_{H}^{\dagger/2} \preceq (1 + \eps) \cdot I_{\mathrm{Im}(L_H)},
 \end{align}
where $I_{\mathrm{Im}(L_H)}$ is simply the projection operator on to $\mathrm{Im}(L_H)$\footnote{This technicality is due to the fact that $L_H$ and $L_H^{\dagger}$ have a non-trivial null-space (i.e., they will always contain the vector of all $1$'s).}. Next, we note that \Cref{eq:pseudoinverseCheck} is true if and only if all non-trivial eigenvalues of $L_{H}^{\dagger/2} \cdot L_{G(Y)} \cdot L_{H}^{\dagger/2}$ are in $(1 \pm \eps)$. This shows that $G(Y)$ is a $(1 \pm \eps)$ spectral sparsifier of $H$ if and only if every (non-trivial) eigenvalue of $L_{H}^{\dagger/2} \cdot L_{G(Y)} \cdot L_{H}^{\dagger/2}$ is in the range $(1 \pm \eps)$. 

Next, we want to show that if we identify an eigenvector $v$ of $L_{H}^{\dagger/2} \cdot L_{G(Y)} \cdot L_{H}^{\dagger/2}$ with eigenvalue $\lambda \notin (1 \pm \eps)$, then we can use this to find a violated constraint in our convex program.

Indeed, let us suppose that we recover such a $v$ and $\lambda$:
 \[
    L_{H}^{\dagger/2} \cdot L_{G(Y)}\cdot L_{H}^{\dagger/2} \cdot v = \lambda \cdot v;
\]
by left multiplying with $v^T$ this yields
\[
    v^\top \cdot  L_{H}^{\dagger/2} \cdot L_{G(Y)}\cdot L_{H}^{\dagger/2} \cdot v = \lambda \cdot v^\top v.
\]
However, for the vector $L_{H}^{\dagger/2} \cdot v$, we also see that \[
v^\top \cdot  L_{H}^{\dagger/2} \cdot L_{H} \cdot L_{H}^{\dagger/2} \cdot v = v^\top \cdot  L_{H}^{\dagger/2} \cdot L_{H}^{1/2} \cdot L_H^{1/2} \cdot L_{H}^{\dagger/2} \cdot v = v^T v.
\]
Thus, if $\lambda \notin (1 \pm \eps)$, we have 
\[
v^\top \cdot  L_{H}^{\dagger/2} \cdot L_{G(Y)}\cdot L_{H}^{\dagger/2} \cdot v = \lambda \cdot v^{\top} v \notin (1 \pm \eps) \cdot v^{\top} v = (1\pm \eps) \cdot v^\top \cdot  L_{H}^{\dagger/2} \cdot L_{H} \cdot L_{H}^{\dagger/2} \cdot v,
\]
and so we can use the constraint specified by $z = L_{H}^{\dagger/2} \cdot v$ as a violated constraint in the convex program. Since calculating pseudo-inverses, multiplying matrices, and finding eigenvalues can all be done in polynomial time, the procedure above gives a polynomial-time separation oracle. 

To conclude, since the feasible region for the above convex program is non-empty, and we have a polynomial time separation oracle, we can find an assignment to the $Y_e$'s which satisfies all of the above constraints in polynomial time by using the ellipsoid method (see \cite[Theorem 6.4.1]{GLS1988}, for instance). For the feasible solution $Y$ found at the end, 
the fractional graph $G(Y)$ defined by the assignment $Y$ is a $(1 \pm \eps)^2$ spectral sparsifier of $G$ (it is a $(1 \pm \eps)$ sparsifier of $H$, which is in turn a $(1 \pm \eps)$ sparsifier of $G$), with the same total weight as the graph $G$. 
\end{proof}

\subsection{Rounding the Fractional Sparsifier}

Finally, in this section we will show how, given a fractional total weight preserving graph $H$, we can round the weights in such a way that we get a simple (unweighted) graph which is a $(1 \pm \eps)$ cut-sparsifier of $H$, while still preserving the total weight exactly. For any edge $e$ in $H$, we will denote by $w_H(e)$ the fractional weight assigned to the edge $e$, that is, $w_H(e) \in [0,1]$. Towards this goal, we establish the following lemma.

\begin{lemma}\label{clm:rounding}
    Let $H$ be a fractional graph whose minimum cut is at least $\lambda \geq 200 \log(n) / \eps^2$, and whose total edge weight sums to an integer. Then, there is a polynomial time randomized rounding scheme which with high probability recovers a simple graph $\widetilde{H}$ which is a total weight preserving $(1 \pm \eps)$ cut sparsifier of $H$. 
\end{lemma}

To prove this lemma, we require the following classic result due to~\cite{Kar93}.  

\begin{proposition}[Karger's Cut-counting Bound~\cite{Kar93}]\label{prop:cut-counting}
     Let $G$ be any (potentially weighted) graph on $n$ vertices with minimum cut size $\lambda(G)$. Then, the number of cuts in $G$ of size $\leq \alpha \cdot \lambda$ is at most $n^{2 \alpha}$.
 \end{proposition}
 
We are now ready to prove~\Cref{clm:rounding}.

\begin{proof}[Proof of~\Cref{clm:rounding}]
    The rounding scheme itself is elementary: we create a simple, unweighted graph $\widetilde{H}$, where for every edge $e \in H$, we keep $e$ independently with probability $w_H(e)$. First, we will show that this procedure recovers a $(1 \pm \eps)$ sparsifier with probability $\geq 1 - 1 /n^{5}$ (although it may not be total weight preserving). 

    Let $t:= 200\log{(n)}/\eps^2$. Because the minimum cut in $H$ is of size $\geq t$, by Karger's cut-counting bound (\Cref{prop:cut-counting}), we know that for any $\alpha \in \Z^+$, the number of cuts of size at most $ \alpha \cdot t$ is bounded by $n^{2 \alpha}$. Now, fix a cut $S$ of size $\in [\alpha \cdot t, 2 \alpha \cdot t)$. It follows from Chernoff bound that 
    \[
        \Pr\paren{\cut{\widetilde{H}}{S} \notin (1 \pm \eps) \cdot \cut{H}{S}} \leq \exp\paren{-\eps^2 \cdot \alpha \cdot t/12} \leq \exp\paren{-\eps^2 \cdot \alpha \cdot \frac{200\log{(n)}}{12\eps^2}} < n^{-10\alpha};
    \]
    (to apply Chernoff bound, we can simply view the weight contributed by each edge as a Bernoulli random variable. The expected weight of a cut under the sampling procedure is exactly equal to its current weight, and the Chernoff bound then follows simply). 
    
    Taking a union bound over all $n^{4\alpha}$ possible cuts, this then yields that every cut of size $\in [\alpha \cdot t, 2\alpha \cdot t)$ is preserved to within a $(1 \pm \eps)$ factor with probability at least $1 - 1  / n^{6 \alpha}$. Finally, integrating over all choices of $\alpha \ge 1$, we get that every cut is preserved to a factor of $(1 \pm \eps)$ with probability at least $ 1 - 1 / n^5$.

    The next step is to show that we can also sample exactly $\sum_{e} w_H(e)$ edges in $\widetilde{H}$ in our randomized rounding approach. For this, we need the following auxiliary claim.

    \begin{claim}\label{clm:poissonBinomialDistribution}
        Let $p_1, \dots p_m$ each be in $[0,1]$, and let $K = \sum_{i = 1}^m p_i$ be an integer. Now, let $X_i = \mathrm{Bern}(p_i)$ and let the $X_i$'s be independently distributed. Then, 
        \[
        \Pr[\sum_{i = 1}^m X_i = K] \geq \frac{1}{m+1}.
        \]
    \end{claim}
    \begin{proof}
        This follows from the fact that the mode of a Poisson binomial distribution is either its mean, or differs from its mean by at most $1$. In particular, in our case when the mean is an integer $(K)$, it must be the case that the mode $\ell = K$ also (see \cite{TT23}, page 2, Darroch's rule for instance). Now, because the support of the distribution has size at most $m+1$ (i.e., $0, \dots m$), it follows that the mode must occur with probability $\geq 1 / (m+1)$, yielding our claim above. \Qed{clm:poissonBinomialDistribution}
        
    \end{proof}
    To conclude the proof of~\Cref{clm:rounding}, 
    we can apply \Cref{clm:poissonBinomialDistribution} to our randomized rounding procedure. We see that with probability $\geq \frac{1}{n^2}$, we will sample \emph{exactly} $\sum_{e \in H} w_H(e)$ edges in $\widetilde{H}$. This is because our edge sampling procedure is exactly a Poisson binomial distribution fitting the form of \Cref{clm:poissonBinomialDistribution}.

    So, we employ the following simple procedure: for $n^3$ rounds, we randomly sample edges in accordance with the above scheme. With probability $1 - (1 - 1/n^2)^{n^3} = 1 - 2^{- \Omega(n)}$, we know that in at least one round, we will recover a graph $\widetilde{H}$ which exactly preserves the total edge mass compared to $H$. Further, by a union bound over all $n^3$ graphs generated in these rounds, we know that with probability $1 - 1 / n^2$ every single graph we generate will be a $(1 \pm \eps)$ cut sparsifier of $H$. Thus, the graph $\widetilde{H}$ that we return is the first graph which preserves the total weight, and it will be a $(1\pm\eps)$ total weight preserving cut sparsifier of $H$ with high probability.  \Qed{clm:rounding}
    
\end{proof}

\subsection{Concluding the Proof of \Cref{thm:sparsifierConstruction}}

Finally, in this section we synthesize our claims to conclude the proof of \Cref{thm:sparsifierConstruction}.

\begin{proof}[Proof of \Cref{thm:sparsifierConstruction}]
    First, we initialize the linear sketch of \Cref{lem:linearSketchParts}, using $\lambda = 200 \log(n) / \eps^2$. The space complexity of our sketch then follows from \Cref{lem:linearSketchParts}. 
    
    Using our sketch, we can recover all edges of strength at most $\lambda$ (we denote this set by $T$), as well as a $(1 \pm \eps)$ spectral sparsifier of the graph $G - T$, and the total number of edges in $G - T$. Next, using \Cref{lem:convexProgram}, we find a fractional $(1 \pm 3\eps)$ total weight preserving spectral sparsifier $H$ of $G - T$ in polynomial time. Finally, we use \Cref{clm:rounding} to, with high probability, round this into a simple $(1 \pm \eps)$ total weight preserving cut sparsifier $\widetilde{H}$ of $H$ in polynomial time. By composition of the sparsifier approximations, we also get that $\widetilde{H}$ is a $(1 \pm 5 \eps)$ simple total weight-preserving cut sparsifier of $G - T$. Finally, we  add back the edges from $T$, and conclude that $\widetilde{H} \cup T$ is a $(1 \pm 5 \eps)$ simple total weight-preserving cut sparsifier of $G$. 
    
    We remark on a minor subtlety here. It seems possible that an edge in $T$ is also included in our rounded solution $\widetilde{H}$, and thus appears twice in $\widetilde{H} \cup T$. However, recall that when we remove edges in $T$, this partitions $G$ into connected components $V_1, \dots V_k$ each with minimum cut greater than 
    $\lambda$. The set of edges $T$ is exactly the set of all edges in $G$ that go across these components. We observe that our spectral sparsifier for the graph $G - T$ will thus not contain any edges crossing $V_1, \dots V_k$. This in turn implies that the fractional graph $H$ generated by our convex program will not have any edges crossing between $V_1, \dots V_k$ also, as otherwise, for some $V_i$, $\cutt_{H}(V_i)$ would be non-zero, whereas $\cutt_{G-T}(V_i) = 0$, violating the spectral approximation constraint in our convex program. So, for every edge $e$ in $T$, $e$ is not present in the rounded graph $\widetilde{H}$. 

    Finally, we can re-parameterize $\eps$ to some $\Theta(\eps)$ to obtain a $(1\pm \eps)$ cut sparsifier with high probability, concluding the proof. 
\end{proof}

Before moving on from this section, we note
that by combining \Cref{lem:convexProgram} and \Cref{clm:rounding}, we get the following general de-sparsification corollary that formalizes part 1 of~\Cref{res:de-sparsify-general}. 

\begin{corollary}[Part 1 of \Cref{res:de-sparsify-general}]\label{lem:generalDesparsification}
     Given a (potentially weighted) graph $H$ which is promised to be a $(1 \pm \eps)$ spectral sparsifier of some unweighted simple graph $G$ with minimum cut $\lambda \geq 200 \log(n) / \eps^2$, along with the number of edges in $G$, we can recover in polynomial time a simple graph $\widetilde{G}$ which is a $(1 \pm O(\eps))$ total weight preserving cut sparsifier of $G$,
     with high probability.
\end{corollary}

%% file: spectral_to_spectral.tex
\section{De-sparsifying Spectral Sparsifiers: Proofs of~\Cref{res:de-sparsify-spectral} and~\ref{res:de-sparsify-general}}\label{sec:res-spectral}

In the previous section, we focused on recovering an unweighted, simple total weight preserving \emph{cut} sparsifiers of our original graph. This was motivated by applications to correlation clustering but also leads to a simpler analysis, as the rounding procedure yields correct outputs so long as the minimum cut value is sufficiently large. We now show that we can also recover a spectral sparsifier of the original graph via de-sparsification. 

\begin{theorem}\label{thm:formalResult2}
    There is a randomized linear sketch of size $\widetilde{O}(n / \eps^2)$ bits which, on any graph $G$, can be used to recover, with high probability, an unweighted, simple $(1 \pm \eps)$ total weight preserving spectral sparsifier of $G$ in polynomial time.
\end{theorem}

To prove~\Cref{thm:formalResult2}, we follow the strategy of~\cite{SS11} in constructing spectral sparsifiers by
sampling edges proportional to their effective resistance (\Cref{def:ER}). Formally, 

\begin{proposition}[\!\!\cite{SS11}]\label{prop:SS11}
    Let $G$ be an arbitrary graph on $n$ vertices, let $\eps \in (0,1)$, and let $C$ be a sufficiently large constant. Then, independently sampling each edge with probability
    \[
    p_e \geq w_e \cdot \frac{C \log(n) \cdot R_{\mathrm{eff}, G}(e)}{\eps^2}
    \]
    (and assigning weight $w_e / p_e$ if sampled) yields a $(1 \pm \eps)$ spectral sparsifier of $G$ with high probability. 
\end{proposition}

\subsection{De-sparsifying With Small Effective Resistances}

In this subsection, we prove part 2 of \Cref{res:de-sparsify-general}, as its intuition will be very valuable in the proof of \Cref{res:de-sparsify-spectral}. We first restate the result before providing our proof:

\begin{theorem}[Part 2 of \Cref{res:de-sparsify-general}]\label{thm:p2-general}
For any $\eps \in (0,1)$ and $n$-vertex unweighted graph $G$, there is a randomized polynomial-time algorithm that given any $(1 \pm \eps)$-spectral sparsifier $H$ of $G$, recovers with high probability
an $n$-vertex, simple unweighted graph $\widetilde{G}$ such that $\widetilde{G}$ is a $(1 \pm 5\eps)$-spectral sparsifier of $G$, provided the effective resistance of every pair $(u,v)$ in $G$ is $\leq\frac{\eps^2}{2C \log(n)}$, where $C$ is the constant from \Cref{prop:SS11}.
\end{theorem}

\begin{proof}
    First, recall that by \cref{lem:convexProgram} we can recover a fractional, total weight preserving graph $\widetilde{H}$ which is a $(1 \pm \eps)$ spectral sparsifier of $H$ and thus a $(1 \pm 3 \eps)$ spectral sparsifier of $G$. All that remains is to perform a randomized rounding of $\widetilde{H}$ which yields a spectral sparsifier and preserves the total weight. We use the same rounding procedure as in \Cref{clm:rounding}, and thus, just as in the proof of the lemma, by repeating the procedure a polynomial number of times, we can guarantee that the number of edges in our rounding matches the number of edges in the original graph. All that remains to be shown is that the rounded graph is a spectral sparsifier with high probability. 

    For this, by \Cref{prop:SS11}, we know that sampling every edge $e$ with probability 
    \[
    p_e \geq \frac{C \log(n)}{\eps^2} \cdot w_e \cdot R_{\mathrm{eff}}(e),
    \] and assigning weight ${w_e}/{p_e}$ to the sampled edges yields a $(1 \pm \eps)$ spectral sparsifier with high probability. By the hypothesis of our theorem, it follows that every edge $e \in \widetilde{H}$ will have effective resistance at most
    \[
    (1 + 2 \eps) \cdot \frac{\eps^2}{2C \log(n)}.
    \]
    This is because $\widetilde{H}$ is a $(1 \pm \eps)$ spectral sparsifier of $G$, and so for every pair of vertices $(u,v) \in \binom{V}{2}$, it is the case that $R_{\mathrm{eff}, \widetilde{H}}(u,v) \in (1 \pm 2\eps)R_{\mathrm{eff}, G}(u,v)$ (see \Cref{def:ER}).
    
    So, by \Cref{prop:SS11} we must only sample each edge $e \in \widetilde{H}$ with probability 
    \[
    p_e \geq (1 + 2 \eps) \cdot \frac{C \log(n)}{\eps^2} \cdot w_e \cdot \frac{\eps^2}{2C \log(n)} = \frac{1+2\eps}{2} \cdot w_e.
    \]
    Since $w_e \geq p_e$, we can keep each edge $e$ independently with probability $w_e$, while still yielding a $(1 \pm \eps)$ spectral sparsifier of $H$ with high probability. Indeed, because $\widetilde{H}$ was already a $(1 \pm 3\eps)$ spectral sparsifier to $G$, the resulting graph is a $(1 \pm 5 \eps)$ spectral sparsifier of $G$ while also being a simple graph.
\end{proof}

\subsection{Proof of \Cref{thm:formalResult2}}

We now provide a formal proof of \Cref{thm:formalResult2}. The main challenge here is to handle the pairs of vertices whose effective resistances will be higher than the bounds in~\Cref{thm:p2-general} which requires a non-black-box modification of our approach in establishing~\Cref{thm:sparsifierConstruction}. 

To prove~\Cref{thm:formalResult2}, we need to use the following more detailed analysis from \cite{KLMMS14}. 

\begin{proposition}[\cite{KLMMS14}]\label{thm:KLMMS}
    Given any parameter $\phi \geq 0$, there is a (randomized) linear sketch $\mathcal{S}$ such that for a graph $G$ on $n$ vertices, $\mathcal{S}(G)$ can be used to recover each edge $e$ independently with probability at least $\phi \cdot r_{\text{eff}, G}(u,v)$, and likewise assigns appropriate weights to create a $(1 \pm \eps)$ spectral sparsifier with probability $1 - 1/ \mathrm{poly}(n)$. Further, $\mathcal{S}$ requires only $\widetilde{O}(n \phi)$ bits of space to store. 
\end{proposition}

We are now ready to start the proof. 

\paragraph{Linear sketch.}
The linear sketch in~\Cref{thm:formalResult2} is very simple: we simply store the sketch of \Cref{thm:KLMMS} with parameter $\phi = \frac{C \log^3(n)}{\eps^2}$, for $C$ a sufficiently large constant. It follows then by \Cref{thm:KLMMS} that this allows us to recover a $(1 \pm \eps)$ spectral sparsifier of the graph $G$ with high probability. Further, since the linear sketch of \Cref{thm:KLMMS} implicitly performs effective-resistance sampling, it follows that every edge $e$ with effective resistance $\geq \frac{1}{\phi} = \frac{\eps^2}{C \log^3(n)}$ is necessarily recovered, as each such edge is sampled with probability $\geq \frac{\phi}{\phi} = 1$.

\paragraph{Recovery from the sketch.}
    Now, let $\widetilde{H}$ denote the weighted spectral sparsifier recovered by the above sketch, and let $\widetilde{H}_U$ denote the corresponding unweighted version of $\widetilde{H}$ (where every edge in $\widetilde{H}$ is given weight $1$). Observe that it must be the case that $\widetilde{H}_U \subseteq G$, as it is the result of sub-sampling $G$. Observe also that because $\widetilde{H}$ is a $(1 \pm \eps)$ spectral sparsifier of $G$, for every pair of vertices $(u,v) \in \binom{V}{2}$, it is the case that $R_{\mathrm{eff}, \widetilde{H}}(u,v) \in (1 \pm 2\eps) R_{\mathrm{eff}, G}(u,v) $ (see \Cref{def:ER}).

    Now, we define the set $\hat{E} \subseteq \binom{V}{2}$:
    \[
    \hat{E} = \left \{(u,v) \in \binom{V}{2}: (u,v) \notin \widetilde{H}_U, R_{\mathrm{eff}, \widetilde{H}}(u,v) \leq \frac{\eps^2}{100 \log^2(n)} \right \}.
    \]
    In words, this is simply the set of pairs of vertices which have small effective resistance with respect to the recovered spectral sparsifier $\widetilde{H}$. Using this we create our convex program, for which we wish to find a feasible point:
\begin{ourbox}
    \begin{align*}
    & Y_e \in [0,1] \quad \forall e \in \hat{E},  \\
    & z^T L_{\widetilde{H}_U} z + \sum_{e \in \binom{V}{2}} Y_e \cdot z^T L_e z \geq (1 - \eps)z^T L_{\widetilde{H}} z \quad \forall z \in \R^n: \Vert z \Vert_2 = 1, \\
    &  z^T L_{\widetilde{H}_U} z + \sum_{e \in \binom{V}{2}} Y_e \cdot z^T L_e z \leq (1 + \eps)z^T L_{\widetilde{H}} z \quad \forall z \in \R^n: \Vert z \Vert_2 = 1, \\
    & |\widetilde{H}_U| + \sum_{e \in \binom{V}{2}} Y_e = m.
\end{align*}
\end{ourbox}

As before, we must show that this program is feasible:

\begin{claim}\label{clm:LPfeasibleSpectral}
    The stated convex program is feasible.
\end{claim}

\begin{proof}
    This follows because the original graph $G$ will be a $(1 \pm \eps)$ spectral sparsifier of $\widetilde{H}$ which preserves the total weight. Because we are already including the contribution of $\widetilde{H}_U$ in each of the constraints, there is a feasible solution corresponding to $G - \widetilde{H}_U$, 
which will contain only edges that are in $\hat{E}$. This is because any edge in $G$ with effective resistance larger than $\frac{\eps^2}{100 \log^2(n)}$ is already in $\widetilde{H}_U$, so the entire support of $G - \widetilde{H}_U$ is thus in $\hat{E}$.
\end{proof}

Likewise, the separation oracle is efficiently implementable:

\begin{claim}\label{clm:LPseparationSpectral}
    There is a polynomial time separation oracle for the above convex program. 
\end{claim}

\begin{proof}
    This follows from all of the same reasons as in \Cref{lem:convexProgram}. Indeed certifying the constraints that $Y_e \in [0,1]$ and that $\sum_{e \in  L_{\widetilde{H}_U}} 1 + \sum_{e \in \binom{V}{2}} Y_e = m$ are both trivial. Thus, it remains only to check whether 
    \[
    z^T L_{\widetilde{H}_U} z + \sum_{e \in \binom{V}{2}} Y_e \cdot z^T L_e z \in (1 + \eps)z^T L_{\widetilde{H}} z \quad \forall z \in \R^n: \Vert z \Vert_2 = 1.
    \]
    Letting $\hat{G}$ denote the graph whose edge weights are given by $Y_e$ (and is $0$ otherwise), this constraint is equivalent to 
    \[
    z^T L_{\widetilde{H}_U} z  + z^T L_{\hat{G}} z \in (1 \pm \eps) z^T L_{\widetilde{H}} z\quad \forall z \in \R^n: \Vert z \Vert_2 = 1.
    \]
    By linearity of the Laplacians, this can be re-written as 
    \[
    z^T (L_{\widetilde{H}_U} + L_{\hat{G}})  z  \in (1 \pm \eps) z^T L_{\widetilde{H}} z\quad \forall z \in \R^n: \Vert z \Vert_2 = 1,
    \]
    which is now exactly in the same form as the constraints of \Cref{lem:convexProgram}, and so can be checked by calculating the eigenvalues of $L_{\widetilde{H}}^{\dagger/2}(L_{\widetilde{H}_U} + L_{\hat{G}})L_{\widetilde{H}}^{\dagger/2} $.
\end{proof}

Now, because of \Cref{clm:LPfeasibleSpectral} and \Cref{clm:LPseparationSpectral}, we can use the ellipsoid method to find a feasible solution in polynomial time \cite{GLS1988}. So, let $\hat{G}$ then denote this feasible solution recovered by the above program, where the edge set is $\hat{E}$, and the corresponding weight on each edge $e \in \hat{E}$ is $Y_e$. Observe that $\hat{G} \cup \widetilde{H}_U$ is a fractional $(1 \pm \eps)$ spectral sparsifier of $\widetilde{H}$, and thus a fractional total weight preserving $(1 \pm 3\eps)$ spectral sparsifier of $G$.

All that remains is to show that efficiently rounding this solution is possible. 
For this, by \Cref{prop:SS11}, we know that sampling each edge $e$ with probability $p_e \geq \frac{C \log(n)}{\eps^2} \cdot w_e \cdot R_{\mathrm{eff}}(e)$, and giving weight $\frac{w_e}{p_e}$ yields a $(1 \pm \eps)$ spectral sparsifier with high probability. Now, because $\hat{G} \cup \widetilde{H}_U$ is a $(1 \pm \eps)$ spectral sparsifier of $\widetilde{H}$, it follows that for every pair of vertices $(u,v)$,
\[
R_{\mathrm{eff}, \hat{G} \cup \widetilde{H}_U}(u,v) \leq (1 + 2\eps)R_{\mathrm{eff}, \widetilde{H}}(u,v).
\]
In particular, for every edge $(u,v) \in \hat{E}$, we have 
\[
R_{\mathrm{eff}, \hat{G} \cup \widetilde{H}_U}(u,v) \leq (1 + 2\eps)\frac{\eps^2}{100 \log^2(n)}.
\]
Thus, in the graph $\hat{G} \cup \widetilde{H}_U$, for every edge $e \in \hat{E}$, we calculate the minimal sampling rate as 
\[
p_e = \frac{C \log(n)}{\eps^2} \cdot w_e \cdot R_{\mathrm{eff}, \hat{G} \cup \widetilde{H}_U}(e) \leq \frac{C \log(n)}{\eps^2} \cdot w_e \cdot \frac{\eps^2}{C \log(n)} \leq w_e.
\]
So, we can independently keep each edge $e \in \hat{G}$ with probability $w_e$ while still creating a $(1 \pm \eps)$ spectral sparsifier of $\hat{G} \cup \widetilde{H}_U$. By composing sparsifiers (as before) this yields a $(1 \pm 5\eps)$ spectral sparsifier of the original graph $G$, while also yielding a simple, unweighted graph (the edges in $\widetilde{H}_U$ are already unweighted). 

By starting with an error parameter of $\eps / 5$, we then obtain the stated accuracy of our spectral sparsifier.
Likewise, because we are performing the simple, independent Bernoulli rounding, we can repeat this procedure $n^3$ times and be guaranteed by \Cref{clm:poissonBinomialDistribution} that in some round, the total weight is exactly preserved. 

This concludes the proof of~\Cref{thm:formalResult2}. \qed

%% file: sublinear_algos.tex
\section{Sublinear Algorithms}\label{sec:sublinear-algs}

In this section, we provide the proofs of \Cref{cor:distributed,cor:mpc,cor:dynamicStreaming,cor:insertionStreaming}. We start with a brief review of the sublinear algorithms models we consider as well as formalizing the role of linear sketching for solving problems in these models. We then provide the 
proofs of the corollaries.

\subsection{Sublinear Algorithms Models}\label{sec:sub-models}

In this paper, we will be working with the following sublinear algorithms models. 

\paragraph{Distributed communication.} In this model, 
the input graph $G=(V,E)$ is edge partitioned across $k$ machines. The machines can communicate with each other in a message-passing model (i.e., the communication is point to point), and the goal is to minimize the total communication between the machines. At the end, one designated machine should output the answer.

\paragraph{Massively Parallel Computation (MPC).} 
In this model, the input graph $G=(V,E)$ is edge partitioned
across multiple machines initially. Computation happens in synchronous rounds wherein each machine can send and receive $\Ot(n)$-size messages. The goal is to have a small number of rounds (ideally a small constant) and compute the final outcome on a designated machine. 

We note that this model is often referred to as near-linear-memory MPC as opposed to the fully scalable MPC wherein the memory of each machine can be any $n^{\delta}$ for  constant $\delta > 0$. 

\paragraph{Dynamic Streams.} In this model, the input graph $G=(V,E)$ is specified as a sequence of insertion and deletion to its edges in a stream. Specifically, each entry of the stream is of the form $(u,v,+)$ or $(u,v,-)$ which either inserts a new edge $(u,v)$ to $G$ or remove an already existing edge $(u,v)$ from $G$, respectively. The algorithm can make one or a few passes over the stream and needs to output the final outcome on the resulting graph at the end of the last pass. 

\subsection{Distributed Communication Model}

We start with proving \Cref{cor:distributed} restated below. 

\begin{corollary*}[Restatement of \Cref{cor:distributed}]
    There is a polynomial-time randomized algorithm for correlation clustering in the distributed communication model with $k$ machines that uses $\Ot(nk)$ communication in total, and with high probability, achieves an $(\alphabest+o(1))$-approximation. 
\end{corollary*}

\begin{proof}
 One of the simplest ways of using linear sketching is to design distributed communication protocols. The coordinator samples the sketching matrix $\mathcal{S}(\cdot)$ and shares it with all the other machines. Then, each machine $i \in [k]$, computes $\mathcal{S}(G_i)$ on its own subgraph $G_i$ and sends it to the coordinator. Finally, the coordinator forms $\mathcal{S}(G) = \mathcal{S}(G_1+\ldots+G_k)$ using the linearity of the sketches.
 We can thus use our~\Cref{thm:formalResult1} 
 and achieve a polynomial time protocol with $\Ot(n)$ communication per machine and so $\Ot(nk)$ communication in total. 
 
The only subtlety here is due to the randomness of the linear sketch and how the sketching matrix $\mathcal{S}$ can be communicated efficiently between the machines. 
Because the linear sketches we are using in \Cref{lem:linearSketchParts} are just the spectral sparsification sketch of \Cref{clm:KLMMS}\cite{KLMMS14} and the spanning forest sketch of \Cref{clm:KPS}\cite{AhnGM12a, KPS24d}, and both these works have already shown how to use only $\widetilde{O}(n)$ many random bits to create the sketch, sharing the sketching matrix is also possible in $\Ot(n)$ communication.
\end{proof}

\subsection{MPC Algorithms}

We next prove \Cref{cor:mpc} restated below. 

\begin{corollary*}[Restatement of \Cref{cor:mpc}]
	There is a polynomial-time randomized algorithm for correlation clustering in the MPC model that uses $O(1)$ rounds and $\Ot(n)$-size messages per machine, and with high probability, achieves an $(\alphabest+o(1))$-approximation. 
\end{corollary*}

To prove this theorem, we need to use the specific form of graph sketching used by our algorithms, referred to as vertex-incidence sketches.

\begin{definition}
    We say a linear sketch $\mathcal{S}(G)$ is a \textbf{vertex-incidence sketch} iff 
    \[
    \mathcal{S}(G) = \Bracket{\mathcal{S}_1(N(v_1)), \mathcal{S}_1(N(v_2)), \dots, \mathcal{S}_n(N(v_n))},
    \]
    where each $\mathcal{S}_i$ is also a linear sketch. Namely, the entire sketch is obtained as a linear sketch of neighborhood of each vertex separately. 
\end{definition}

The following is a well-known fact about MPC protocols for combining vertex-incidence sketches.

\begin{proposition}[\cite{AhnGM12b, AKLP22}]\label{prop:AKLP}
    Let $\mathcal{S}(G)$ denote a vertex-incidence linear sketch which requires $\leq s$ bits of space for each vertex-neighborhood sketch. Then, there is a $2$ round, $\widetilde{O}(n \cdot s)$ communication MPC protocol which results in a single machine containing the entire sketch $\mathcal{S}(G)$.
\end{proposition}

This will be useful for us as the linear sketches of \Cref{clm:KPS} and \Cref{clm:KLMMS} used in our~\Cref{lem:linearSketchParts} are vertex-incidence sketches.

\begin{proof}[Proof of \Cref{cor:mpc}]
Because \Cref{clm:KPS} and \Cref{clm:KLMMS} are both vertex-incidence sketches that require $O(\mathrm{polylog}(n) / \eps^2)$ bits for each vertex-neighborhood, this implies a simple $\widetilde{O}(n / \eps^2)$-size message MPC protocol which terminates with a single machine $M_1$ containing $S_1(G), S_2(G)$, and $S_3(G)$ as in \Cref{lem:linearSketchParts}.

From here, we must only set $\eps = 1 / \log(n)$ and have $M_1$ run the recovery algorithm from \Cref{thm:formalResult1} to conclude our corollary.
\end{proof}

\subsection{Dynamic Streaming}

Next, we provide the proof of \Cref{cor:dynamicStreaming}. 

\begin{corollary*}[Restatement of \Cref{cor:dynamicStreaming}]
	There is a polynomial-time randomized streaming algorithm for correlation clustering that uses $\Ot(n)$ memory when making a single pass over a dynamic stream, and for any constant $\eps > 0$, with high probability, achieves an $(\alphabest+o(1))$-
	approximation. 
\end{corollary*}

\begin{proof}
Another very simple application of linear sketching is to dynamic streaming algorithms. At the beginning of the stream, the algorithm can sample the sketching matrix $\mathcal{S}$ and compute $\mathcal{S}(\emptyset)$ as the sketch. Then, whenever a 
new edge $e$ is inserted or deleted, 
the algorithm can update the sketch by $\mathcal{S}(+e)$ or
$\mathcal{S}(-e)$ using the linearity of the sketch. This way, at the end of the stream, the algorithm is left with a sketch of the final graph. 

In our context, we simply run this approach using our sketching algorithm in~\Cref{thm:formalResult1} and at the end, in polynomial time, return the solution. We note that similar to~\Cref{cor:distributed}, here also, we additionally exploit the fact that the description of the sketching matrix can be stored in $\Ot(n)$ bits.
\end{proof}

\subsection{Deterministic Algorithms for Insertion-Only Streams}

Finally, we show that by leveraging known results on \emph{deterministic} algorithms for spectral sparsification in insertion-only streams, we can achieve the following result. 

\begin{corollary*}[Restatement of \Cref{cor:insertionStreaming}]
	There is a polynomial-time streaming algorithm for correlation clustering that uses $\Ot(n)$ memory to deterministically build a data structure $D$ using a single pass over an insertion-only stream, and only at the end, uses randomization to, with high probability, recover from $D$ an $(\alphabest+o(1))$-approximation for correlation clustering. 
\end{corollary*}

\begin{remark}
    Note that the high probability in~\Cref{cor:insertionStreaming} is with respect to the post-processing, not the data structure $D$ created during the stream itself, which is deterministic. Moreover because the algorithm is deterministic during the stream, it works even against an adversary that sees the internal state of the algorithm, namely, is adversarially robust in the strongest possible sense.  
\end{remark}

The key to the proof of~\Cref{cor:insertionStreaming} is the following lemma. 

\begin{lemma}\label{thm:deterministicSpectralDesparsification}
    There is a polynomial time streaming algorithm that uses $\widetilde{O}(n/\eps^2)$ bits of memory  to deterministically build a data structure $D$ using a single pass over an insertion-only stream of edges of $G$, and, only at the end, uses randomization to recover a simple graph $H$ from $D$ which is a $(1 \pm \eps)$ total weight preserving spectral sparsifier of $G$ with high probability. 
\end{lemma}

 Before proving this lemma, we show how it implies~\Cref{cor:insertionStreaming}. 

\begin{proof}[Proof of \Cref{cor:insertionStreaming}]
   We just run \Cref{thm:deterministicSpectralDesparsification}. The space required is only $\widetilde{O}(n / \eps^2)$ bits. 

    By the statement of \Cref{thm:deterministicSpectralDesparsification}, this yields a simple graph $H$ which is a $(1 \pm \eps)$ total weight preserving spectral sparsifier of $G$, and in particular, also a $(1 \pm \eps)$ cut sparsifier of $G$. 

    Now, on $H$, by \Cref{clm:preserveCCperfectDegrees}, we see that for any clustering $V_1, \dots V_k$, 
    \[
    \mathrm{CC}_{H}(V_1, \dots V_k) \in (1 \pm 2\eps) \mathrm{CC}_{G}(V_1, \dots V_k).
    \]
    So, if we let $\mathrm{OPT}$ denote the minimum correlation clustering value, we know that 
    \[
    \mathrm{OPT}(H) \leq (1 + 2 \eps) \cdot \mathrm{OPT}(G).
    \]

    Now, let us run the $\alphabest$-approximation, polynomial time algorithm for correlation clustering on $H$ (here, we are using the fact that $H$ is simple). We are guaranteed that this recovers a partition $\hat{P}$ such that 
    \[
    \mathrm{CC}_{H}(\hat{P}) \leq \alphabest \cdot \mathrm{OPT}(H).
    \]

    Finally, we note that the solution we recover satisfies
    \[
     \mathrm{CC}_{G}(\hat{P}) \leq (1 + 2 \eps) \mathrm{CC}_{H}(\hat{P}) \leq (1 + 2 \eps) \cdot \alphabest \cdot \mathrm{OPT}(H) \leq \alphabest \cdot (1 + 2 \eps)^2 \cdot \mathrm{OPT}(G).
    \]

    By setting $\eps = 1 / \log(n)$, the total space required is only $\widetilde{O}(n)$ bits, yet still recovers an $(\alphabest + o(1))$-approximate solution to correlation clustering. 
\end{proof}

In the rest of this section, we focus on a proof of \Cref{thm:deterministicSpectralDesparsification}.
The key building block here will be the existence of deterministic spectral sparsifiers of \cite{BSS09}. It is also known that these can be leveraged into deterministic insertion-only spectral sparsification algorithms as follows. 

\begin{proposition}[cf.~\cite{McG14}]\label{clm:insertionOnlySparsifier}
There is a deterministic single-pass streaming algorithm on insertion-only streams that computes a $(1\pm \eps)$ spectral sparsifier using $\Ot(n/\eps^2)$ space. 
\end{proposition}

Note that it is tempting to try to invoke the same reasoning that was used in the proof of \Cref{thm:formalResult2} to argue that the deterministic spectral sparsification algorithm recovers edges with large effective resistance. Unfortunately however, this is not necessarily true with the algorithm of \Cref{clm:insertionOnlySparsifier}. Instead, our algorithm needs to explicitly recover these edges. To do this we first introduce a few definitions:

\begin{definition}
    For a graph $G$, we say that a subgraph $T \subseteq G$ is a $\log(n)$-\textbf{spanner} of $G$, if for every $u,v$:
    \[
    \mathsf{dist}_T(u,v) \leq \log(n) \cdot \mathsf{dist}_G(u,v) ,
    \]
    where $\mathsf{dist}_G(u,v)$ refers to the length of the shortest path between $(u,v)$ in $G$. 
\end{definition}

\begin{definition}
    We say that $T_1, \dots T_{\ell}$ are \textbf{$\ell$ form a sequence of disjoint $\log(n)$-spanners} of $G$ if $\forall i \in [\ell]$, $T_i$ is a $\log(n)$-spanner of the graph $G - T_1 - T_2 - \dots - T_{i-1}$.
\end{definition}

We also rely on the following two propositions, one detailing a standard process for creating $\log(n)$-spanners, and another connecting $\log(n)$-spanners to effective resistance.

\begin{proposition}[cf.~\cite{ABSHJKS20}]\label{prop:ABSH}
    Let $G$ be a graph, and let $T$ be the result of iteratively removing an arbitrary single edge from any cycle of length at least $\log(n)$ that remains in $G$. Then, $T$ is an $\log(n)$-spanner of $G$, and $T$ has only $O(n)$ edges.
\end{proposition}

The work of \cite{ADKKP16} provided the following characterization of effective resistance:

\begin{proposition}[\!\cite{ADKKP16}]\label{clm:spannerER}
Let $G$ be a simple graph, and let $T_1, \dots T_{\ell}$ be $\ell$ be a sequence of disjoint $\log(n)$ spanners. Then, for any edge $e=(u,v) \in G - T_1 - \dots - T_{\ell}$, the effective resistance between $u$ and $v$ is at most $\frac{\log(n)}{\ell}$ in the graph $T_1 \cup T_2 \cup \dots \cup T_{\ell}$. 
\end{proposition}

Using this, we can proceed to the proof of \Cref{thm:deterministicSpectralDesparsification}:

\begin{proof}[Proof of \Cref{thm:deterministicSpectralDesparsification}]
Our data structure $D$ will consist of a sequence of disjoint $\log(n)$-spanners as well as  a deterministic spectral sparsifier of the remaining edges. Specifically, we create a sequence of disjoint $\log(n)$-spanners by processing edges as they arrive in the stream. This procedure is straightforward: whenever a new edge $e$ arrives, we attempt to insert $e$ into the first spanner $T_1$. If this creates a cycle of length $\leq \log(n)$ in $T_1$, then we remove $e$ from $T_1$ and instead try inserting it into $T_2$, and so on. If $e$ does not get included into $T_1, \dots T_{\ell}$, then we insert it into the algorithm of \Cref{clm:insertionOnlySparsifier}.

    Since this process is exactly implementing \Cref{prop:ABSH}, each $T_i$ is a $\log(n)$-spanner of $G - T_1 - \dots T_{i-1}$ (again, see \cite{ABSHJKS20} for a discussion). In particular, this means that every edge $e \in G - \bigcup_{i = 1}^{\ell} T_i$ has effective resistance $\leq \frac{\log(n)}{\ell}$ in $G$ by \Cref{clm:spannerER}. 

    At this point, the algorithm has recovered the sequence of $\ell$ spanners $T_1, \dots T_{\ell}$, as well as a $(1 \pm \eps)$ spectral sparsifier of $G - \bigcup_{i = 1}^{\ell} T_i$, which we denote by $\widetilde{G-T}$. However, observe that we still cannot use the convex program of \Cref{lem:convexProgram}. Although we are guaranteed that every edge in $G - \bigcup_{i = 1}^{\ell} T_i$ has small effective resistance, if we just run the original convex program, it is possible that it assigns fractional mass to edges which are not in the original graph $G$, and therefore have large effective resistances (and hence the randomized rounding scheme would not work). Instead, given the spanners $T_1, \dots T_{\ell}$, we define a new set $\hat{E} \subseteq \binom{V}{2}$ as
    \[
    \hat{E} = \left \{e \in \binom{V}{2}: e \notin \bigcup_{i = 1}^{\ell} T_i \wedge R_{\mathrm{eff}, (\bigcup_{i = 1}^{\ell} T_i)}(e) \leq \frac{\log(n)}{\ell} \right \}.
    \]

    Here, we use $R_{\mathrm{eff}, (\bigcup_{i = 1}^{\ell} T_i)}(e)$ to denote the effective resistance of an edge $e$ in the graph formed only by the edges contained in the sequence of spanners $T_1, \dots T_{\ell}$. We can now introduce a modified convex program whose solution support is guaranteed to be in $\hat{E}$ while  remaining feasible:
\begin{ourbox}
\begin{align*}
    & Y_e \in [0,1] \quad \forall e \in \hat{E},  \\
    & \sum_{e \in \binom{V}{2}} Y_e \cdot z^T L_e z \geq (1 - \eps)z^T L_{\widetilde{G-T}} z \quad \forall z \in \R^n: \Vert z \Vert_2 = 1, \\
    & \sum_{e \in \binom{V}{2}} Y_e \cdot z^T L_e z \leq (1 + \eps)z^T L_{\widetilde{G-T}} z \quad \forall z \in \R^n: \Vert z \Vert_2 = 1, \\
    & \sum_{e \in \binom{V}{2}} Y_e = m - \sum_{i = 1}^{\ell}|T_i|.
\end{align*}
\end{ourbox}

\begin{claim}\label{clm:LPfeasibleSpectral2}
    The convex program above is feasible.
\end{claim}

\begin{proof}
This follows because all edges $G - \bigcup_{i = 1}^{\ell} T_i$ are in $\hat{E}$ (\Cref{clm:spannerER}), and hence constitute a $(1 \pm \eps)$ spectral sparsifier of $\widetilde{G-T}$. Moreover, this solution also preserves the total weight, and only uses $\{0,1\}$ valued edge weights.
\end{proof}

We next observe that the separation oracle for the above convex program is efficiently implementable:

\begin{claim}\label{clm:LPseparationSpectral2}
    There is a polynomial-time separation oracle for the above convex program.
\end{claim}

\begin{proof}
    This follows from essentially same reasoning as done in \Cref{lem:convexProgram}. Indeed certifying the constraints that $Y_e \in [0,1]$ and that $\sum_{e \in \binom{V}{2}} Y_e = m - \sum_{i = 1}^{\ell}|T_i|$ are both trivial. Thus, it remains only to check the spectral approximation conditions. 

    This can be done by checking the eigenvalues of $L_{\widetilde{G-T}}^{\dagger/2} L_{\hat{G}} L_{\widetilde{G-T}}^{\dagger/2}$ using the same logic as \Cref{lem:convexProgram} and \Cref{clm:LPseparationSpectral}.
\end{proof}

Now, because of \Cref{clm:LPfeasibleSpectral2} and \Cref{clm:LPseparationSpectral2}, we can use the ellipsoid method to find a feasible solution in polynomial-time \cite{GLS1988}. So, let $\hat{G}$ then denote this feasible solution recovered by the above program, where the edge set is $\hat{E}$, and the corresponding weight on each edge $e \in \hat{E}$ is $Y_e$. Observe that $\hat{G} \cup \bigcup_{i = 1}^{\ell} T_i$ is a fractional $(1 \pm \eps)$ spectral sparsifier of $\widetilde{G-T} \cup \bigcup_{i = 1}^{\ell} T_i$, and thus a fractional total weight preserving $(1 \pm 3\eps)$ spectral sparsifier of $G$.

All that remains is to show that efficiently rounding our fractional $(1 \pm \eps)$ spectral sparsifier $\hat{G} \cup \bigcup_{i = 1}^{\ell} T_i$ is possible. By \Cref{prop:SS11}, sampling every edge $e$ in $\hat{G} \cup \bigcup_{i = 1}^{\ell} T_i$ with probability $p_e \geq \frac{C \log(n)}{\eps^2} \cdot w_e \cdot R_{\mathrm{eff}, \hat{G} \cup \bigcup_{i = 1}^{\ell} T_i}(e)$, and giving weight $\frac{w_e}{p_e}$ yields a $(1 \pm \eps)$ spectral-sparsifier with high probability. 

By setting $\ell = \frac{C \log^2(n)}{\eps^2}$ (using the constant $C$ in \Cref{prop:SS11}), it follows that every edge $e \in \hat{E}$ will have effective resistance $\leq \frac{\log(n)\eps^2}{C \log^2(n)} = \frac{\eps^2}{C \log(n)}$ in $\bigcup_{i = 1}^{\ell} T_i$.
Because effective resistance only decreases as one adds more edges, it also follows that every edge $e \in \hat{E}$ will have effective resistance $\leq \frac{\eps^2}{C \log(n)}$ in $\hat{G} \cup \bigcup_{i = 1}^{\ell} T_i$. 
Thus, in the graph $\hat{G} \cup \bigcup_{i = 1}^{\ell} T_i$, if we keep every edge $\bigcup_{i = 1}^{\ell} T_i$ with probability $1$, and then sample every edge $e \in \hat{G}$ with probability $\frac{C \log(n)}{\eps^2} \cdot w_e \cdot \frac{\eps^2}{C \log(n)} = w_e$, we will get with high probability a $(1 \pm \eps)$ spectral sparsifier of $\hat{G} \cup \bigcup_{i = 1}^{\ell} T_i$. We denote this resulting simple graph by $H$. Now since $\hat{G} \cup \bigcup_{i = 1}^{\ell} T_i$ is a $(1 \pm 3 \eps)$ spectral sparsifier to $G$, it follows that $H$ is a $(1 \pm 5\eps)$ spectral sparsifier of $G$. Thus running the above procedure with an error parameter of $\eps / 5$ yields a spectral sparsifier with desired accuracy.

Finally, because we are performing the simple, independent Bernoulli rounding, we can repeat this procedure $n^3$ times and be guaranteed by \Cref{clm:poissonBinomialDistribution} that in some round, the total weight is exactly preserved. The memory required by the algorithm follows from the $\widetilde{O}(n / \eps^2)$ edges stored in the spanners (by our choice of $\ell$ and \Cref{prop:ABSH}), and the complexity of the deterministic spectral sparsifier (\Cref{clm:insertionOnlySparsifier}). This yields the lemma. 
\end{proof}

%% file: appendix.tex
\section{Omitted Proofs}\label{app:omitted}

\subsection{Proof of~\Cref{clm:preserveCCperfectDegrees}}\label{sec:clm:preserveCCperfectDegrees}
We now provide a proof of \Cref{clm:preserveCCperfectDegrees}, restated below. 

\begin{lemma*}[\Cref{clm:preserveCCperfectDegrees}]
    Let $G$ and $H$ be graphs on the same vertex set such that $H$ is a $(1 \pm \eps)$ total weight preserving cut sparsifier of $G$. Then, for any partition $V_1, \dots V_k$ of vertices,
    \[
    \mathrm{CC}_H(V_1, \dots V_k) \in 
    (1 \pm 2\eps) \cdot \mathrm{CC}_G(V_1, \dots V_k).
    \]
\end{lemma*}
\begin{proof}
Recall that by~\Cref{def:cc},
\[
\mathrm{CC}_G(V_1, \dots V_k) = \sum_{i \in [k]} |E^-(V_i)| + |E^+(V_1, \dots V_k)|,
\]
Observe that we can re-write $|E^+(V_1, \dots V_k)|$ as
\[
|E^+(V_1, \dots V_k)| = \frac{1}{2} \cdot \sum_{i \in [k]} \cutt_G(V_i),
\]
as each crossing edge will be present in exactly two of the $V_i$ cuts. Similarly, 
\[
|E^-(V_i)| = \binom{|V_i|}{2} - \frac{1}{2} \left (\sum_{v \in V_i} \deg(v) - \cutt_G(V_i) \right ),
\]
as the second term above counts the number of edges inside each $V_i$. Together, these mean 
\[
\mathrm{CC}_G(V_1, \dots V_k) = \sum_{i \in [k]}\cutt_G(V_i) + \sum_{i \in [k]} \binom{|V_i|}{2} - \frac{1}{2}\sum_{v \in V} \deg(v).
\]
\noindent
Moreover, since $G$ and $H$ have the same total weight, we have
\[
\sum_{i \in [k]} \binom{|V_i|}{2} - \frac{1}{2}\sum_{v \in V} \deg_H(v) = \sum_{i \in [k]} \binom{|V_i|}{2} - \frac{1}{2}\sum_{v \in V} \deg_G(v). 
\]    
Thus, the only difference between $\mathrm{CC}_G$ and $\mathrm{CC}_H$ is in the terms $\sum_{i \in [k]}\cutt_G(V_i)$ vs. $\sum_{i \in [k]}\cutt_H(V_i)$. Because $H$ is a $(1 \pm \eps)$ cut sparsifier of $G$, it follows that 
\[
    \left | \mathrm{CC}_H(V_1, \dots V_k) - \mathrm{CC}_G(V_1, \dots V_k) \right | 
    = \left |\sum_{i \in [k]}\cutt_H(V_i) - \sum_{i \in [k]}\cutt_G(V_i)\right |
    \leq \eps \cdot \sum_{i \in [k]}\cutt_G(V_i).
\]
Finally, we also have
\[
\mathrm{CC}_G(V_1, \dots V_k) \geq \sum_{i \in [k]} |E^+(V_1, \dots V_k)| = \frac{1}{2} \cdot \sum_{i \in [k]} \cutt_G(V_i), 
\]
and combining the previous two equations gives us, 
\[
    \left | \mathrm{CC}_H(V_1, \dots V_k) - \mathrm{CC}_G(V_1, \dots V_k) \right | \leq 2 \eps \cdot \mathrm{CC}_G(V_1, \dots V_k),
 \]
 concluding the proof. 
\end{proof}

\subsection{\textbf{NP}-Hardness of Certifying Cut Sparsification}\label{sec:NP}

As stated in~\Cref{sec:intro}, it was important for us to work with spectral sparsifiers in our de-sparsification paradigm even though for our application, cut sparsifiers would have sufficed also. Intuitively, the reason for this is that it is simple to \emph{certify} that two graphs $G_1, G_2$ are spectral sparsifiers of one another. Naturally then, one may wonder if the same is true about certifying whether two graphs $G_1, G_2$ are \emph{cut} sparsifiers of one another. We now show that this is not true. Indeed, we show that being able to certify that arbitrary graphs are $(1 \pm \eps)$ cut sparsifiers of each other is \textbf{NP}-hard in general. 

As a starting point, we recall the following about the sparsest cut problem:

\begin{definition}
    For an arbitrary graph $G = (V, E)$, and a subset $S \subseteq V$, we say that the sparsity of the cut $S$ is
    \[
    \Phi(S) = \frac{|E[S, \bar{S}]|}{|S| |\bar{S}|}.
    \]
    The sparsity of the graph is defined as \[
    \Phi(G) = \min_{S \neq \emptyset, V}\Phi(S).
    \]
\end{definition}

We will rely on the following result:

\begin{proposition}[\!\!\cite{BBPP12}]
    It is $\mathbf{NP}$-hard to calculate $\Phi(G)$ on unweighted simple graphs $G$.
\end{proposition}

Using this we prove the following result (as stated in~\Cref{sec:intro}, we believe this result is folklore but we know no reference for it and as such are proving it here for completeness). 

\begin{proposition}
    It is $\mathbf{NP}$-hard to check if two graphs are $(1 \pm \eps)$ cut sparsifiers of one another.
\end{proposition}

\begin{proof}
    Let $G$ be an arbitrary simple graph for which we wish to approximate $\Phi(G)$.
    
    Next, let us consider the complete graph $K_n$. In order for our graph $G$ to be a $(1 \pm \eps)$ cut sparsifier of $K_n$ it must be the case that for every $S \subseteq V$
    \[
   (1 - \eps) \cutt_{K_n}(S)  \leq \cutt_{G}(S) \leq (1 + \eps) \cutt_{K_n}(S).
    \]
    In particular however, we know that $\cutt_{K_n}(S) = |S||\bar{S}|$, and further, $\cutt_{G}(S) \leq (1 + \eps) \cutt_{K_n}(S)$ by definition, as $K_n$ contains all edges (and more) of $G$. Thus, our graph $G$ is a cut sparsifier of $K_n$ if and only if for every $S \subseteq V$:
    \[
 \cutt_{G}(S) \geq (1 - \eps) |S||\bar{S}|,
    \]
    equivalently, if and only if 
    \[
    \Phi(S) = \frac{|E[S, \bar{S}]|}{|S||\bar S|} \geq (1 - \eps).
    \]

    Now, suppose we have an algorithm which efficiently certifies whether $G$ is a $(1 \pm \eps)$ cut sparsifier of $K_n$ for any value of $\eps$, and let us denote this algorithm by $A(G, K_n, \eps)$. To start, we can query with $\eps = 1 - \frac{1}{n^2}$. This will always return $1$, as we are simply certifying that 
    \[
    \cutt_{G}(S) \geq \frac{1}{n^2} |S||\bar{S}|,
    \]
    which is equivalent to checking that $G$ is connected. Now, our goal is to calculate $\Phi(G)$, or equivalently, to find $\min_{S \neq \emptyset, V} \frac{|E[S, \bar{S}]|}{|S||\bar{S}|}$. Observe that this expression can only take on $\leq n^3$ values, as the number of edges in $E[S,\bar{S}]$ is $\leq n^2$, and the denominator can only take on $n/2$ values (one for each value of $|S| = 1, \dots n/2$). So, there exists a set of $\leq n^3$ possible values for $\Phi(S)$, which we denote by $Q$. Our goal will be to find the smallest value $q \in Q$ for which there exists an $S$ such that $\Phi(S) = q$. 
    
    So, let us sort the set of possible values $Q$, and denote the possible values by $q_1, \dots q_{\leq n^3}$. To start, we set $\eps = (1 - q_2)$, and query $A(G, K_n, \eps)$. Then, we are certifying whether $\forall S \subseteq V$,
\[
    \Phi(S) \geq q_2.
    \]
    If this fails, then there must be a cut with smaller sparsity, and hence $\Phi(G) = q_1$.
    Otherwise, we instead query with $\eps = 1 - q_3$, then $\eps = 1 - q_4$, $\eps = 1 - q_5$, etc. until we find a value $\eps = 1 - q_i$ such that $G$ is not a $(1 \pm \eps)$ cut sparsifier of $K_n$. For this value, we know that $\forall S$,
    \[
     \Phi(S) \geq q_{i-1},
    \]
    but that there exists an $S$ such that 
    \[
    \Phi(S) < q_i.
    \]
    Thus, $\Phi(G)$ will be exactly $q_{i-1}$. This requires only $\mathrm{poly}(n)$ oracle calls to the cut-certification algorithm $A$. Thus, calculating the sparsest cut efficiently reduces to cut certification, meaning the latter must be \textbf{NP}-hard.
\end{proof}